%%%%%%%%%%%%%%%%%%%%%%%%%%%%%%%%%%%%%%%%%%%%%%%%%%%%%%%%%%%%%%%%%%%%%%%%%%
%                                                                        %
%                                                                        % 
%        by P. Di Francesco, E. Guitter and C. Kristjansen               %
%                TEX file, using harvmac.tex macros                      %
%                                                                        %
%                                                                        %
%%%%%%%%%%%%%%%%%%%%%%%%%%%%%%%%%%%%%%%%%%%%%%%%%%%%%%%%%%%%%%%%%%%%%%%%%%
\input harvmac 
\input epsf.tex

\overfullrule=0mm

\newcount\figno
\figno=0
\def\fig#1#2#3{
\par\begingroup\parindent=0pt\leftskip=1cm\rightskip=1cm\parindent=0pt
\baselineskip=11pt
\global\advance\figno by 1
\midinsert
\epsfxsize=#3
\centerline{\epsfbox{#2}}
\vskip 12pt
{\bf Fig. \the\figno:} #1\par
\endinsert\endgroup\par
}
\def\figlabel#1{\xdef#1{\the\figno}}
\def\encadremath#1{\vbox{\hrule\hbox{\vrule\kern8pt\vbox{\kern8pt
\hbox{$\displaystyle #1$}\kern8pt}
\kern8pt\vrule}\hrule}}

%Macros 
%%%%%%%%%%%%%%%%%%%%%%%%%%%%%%%%%%%%%%%%%%%%%%%%%%%%%%%%%%%%%%%%%

\def\IR{\relax{\rm I\kern-.18em R}}
\font\cmss=cmss10 \font\cmsss=cmss10 at 7pt

\font\cmss=cmss10 \font\cmsss=cmss10 at 7pt
\def\IZ{\relax\ifmmode\mathchoice
{\hbox{\cmss Z\kern-.4em Z}}{\hbox{\cmss Z\kern-.4em Z}}
{\lower.9pt\hbox{\cmsss Z\kern-.4em Z}}
{\lower1.2pt\hbox{\cmsss Z\kern-.4em Z}}\else{\cmss Z\kern-.4em Z}\fi}
\def\IN{\relax{\rm I\kern-.18em N}}

%%%%%%%%%%%%%%%%%%%%%%%%%%%%%%%%%%%%%%%%%%%%%%%%%%%%%%%%%%%%%%%%%

\Title{\vbox{\hsize=3.truecm \hbox{SPhT/00-145}\hbox{NBI-HE-00-39}}}
{{\vbox {
%\centerline{}
\bigskip
\centerline{Generalized Lorentzian Triangulations}
\centerline{and the Calogero Hamiltonian}
}}}
\bigskip
\centerline{P. Di Francesco\foot{philippe@spht.saclay.cea.fr},
E. Guitter\foot{guitter@spht.saclay.cea.fr}}
\medskip
\centerline{ \it CEA-Saclay, Service de Physique Th\'eorique,}
\centerline{ \it F-91191 Gif sur Yvette Cedex, France}
\medskip
\centerline{C. Kristjansen\foot{kristjan@nbi.dk, supported by the
Carlsberg Foundation}}
\medskip
\centerline{ \it The Niels Bohr Institute, Blegdamsvej 17,}
\centerline{ \it DK-2100 Copenhagen \O, Denmark} 

\vskip .5in

%abstract
\noindent
We introduce and solve a generalized model of 1+1D Lorentzian 
triangulations in
which a certain subclass of outgrowths is allowed, the occurrence
of these being governed by a coupling constant $\beta$. Combining
transfer matrix-, saddle point- and path integral techniques we show
that for $\beta<1$ it is possible to take a continuum limit in which
the model is described by a 1D quantum Calogero Hamiltonian. 
The coupling constant $\beta$ survives the continuum
limit and appears as a parameter of the Calogero potential.

%\draft
\noindent
\Date{10/00}

%references
\nref\ADJ{J.\ Ambj\o rn, B.\ Durhuus and T.\ Johnsson, {\it Quantum
Geometry}, Cambridge University Press, 1997.}
\nref\canonical{F.\ David, Nucl.\ Phys.\ B257 (1985) 45; V.A.\
Kazakov, Phys.\ Lett.\ 150B (1986) 140.}
\nref\AL{J.\ Ambj\o rn and R.\ Loll, Nucl.\ Phys.\ B536 (1998) 407, 
hep-th/9805108.}
\nref\Nak{R.\ Nakayama, Phys.\ Lett.\ B 325 (1994) 347.}
\nref\DK{B.\ Duplantier and I.\ Kostov, Phys.\ Rev.\ Lett.\ 61 (1988) 559.}
\nref\ACKL{J.\ Ambj\o rn, J.\ Correia, C.\ Kristjansen and R.\ Loll,
Phys.\ Lett.\ B475 (2000) 24, hep-th/9912267.}
\nref\cycles{B.\ Duplantier and I.\ Kostov, Nucl.\ Phys.\ B340 (1990) 491;
B.\ Eynard, E.\ Guitter and C.\ Kristjansen, Nucl.\ Phys.\ B528 (1998) 523,
cond-mat/9801281;
S.\ Higuchi, Mod.\ Phys.\ Lett.\ A 13 (1998) 727, cond-mat/9806349.} 
\nref\AAL{J.\ Ambj\o rn, K.N.\ Anagnostopoulos and R. Loll, 
Phys.\ Rev.\ D60 (1999) 104035,
hep-th/9904012; Phys.\ Rev.\ D61 (2000) 044010, hep-lat/9909129.}
\nref\FGK{P.\ Di Francesco, E.\ Guitter and C.\ Kristjansen, Nucl.\ Phys.\ 
B567 (2000) 515, hep-th/9907084.}
\nref\KPZ{V.G. Knizhnik, A.M. Polyakov and A.B. Zamolodchikov, Mod. Phys. Lett.
{\bf A3} (1988) 819.}
\nref\LL{L.D.\ Landau and E.M.\ Lifshitz,{\it Quantum Mechanics},
Pergamon Press, 1958.}
\nref\BATI{ H. Bateman, {\it Higher Transcendental Functions}, Vol. I,
McGraw-Hill (1953).}
\nref\BAT{H. Bateman, {\it Higher Transcendental Functions}, Vol. II,
McGraw-Hill (1953).}
\nref\ID{C.\ Itzykson and J.-M.\ Drouffe, {\it Statistical Field
Theory}, Cambridge University Press 1989.}
\nref\SPO{H. Spohn, {\it Fixed points of a functional renormalization group
for critical wetting}, Europhys. Lett. {\bf 14} (1991) 689.}

%text

\newsec{Introduction}
Random triangulations are interesting both from a field theoretic and 
from a statistical mechanical point of view (for a review, see
for instance~[\xref\ADJ-\xref\canonical]). In particular, the so-called
dynamical triangulations, also denoted in the following as Euclidean 
triangulations, provide us with a consistent lattice regularized version 
of 2D Euclidean quantum gravity, which in its scaling limit reproduces
the continuum Liouville field theory. Moreover, the introduction of 
statistical mechanical degrees of freedom on dynamical
triangulations also led to the discovery of new universality classes
describing critical phenomena on fluctuating surfaces.

Beside dynamical triangulations, another type of random triangulations 
has recently been introduced, known as 1+1D Lorentzian triangulations~\AL.
What characterizes a Lorentzian triangulation is that it can
be decomposed into slices by cutting along horizontal lines, as
shown in Fig.~1. 
Each slice is made of an arbitrary sequence of triangles pointing up or down at
random. Each triangle has two ``time-like" edges within the slice and one
``space-like" edge shared with a triangle in a neighboring slice. 
\fig{A Lorentzian triangulation (a) and its dual
(b).}{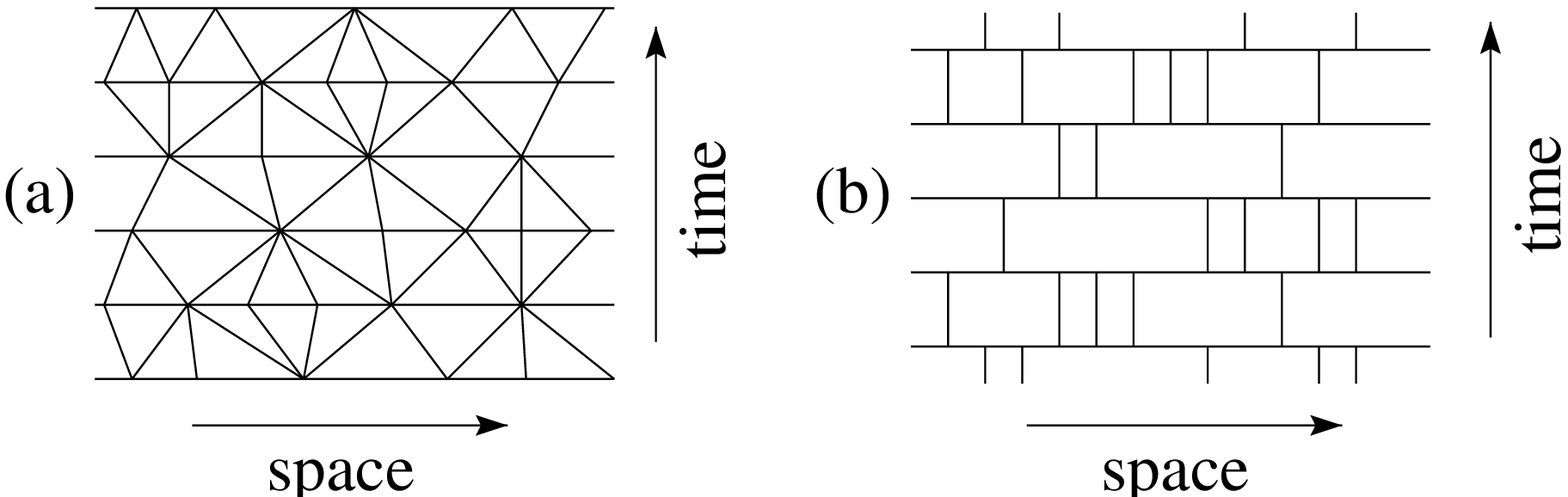}{10.cm}
\figlabel\lorgra
 Lorentzian
triangulations were invented with the aim of constructing a  model
 of quantum
gravity where causality was built in at a fundamental level. 
The resulting gravity model was denoted as Lorentzian gravity~\AL. This model
has a well defined scaling limit which is different from that of Euclidean
gravity. At present there does not exist any continuum formulation 
\`{a} la Liouville with such a causal structure. 
We note, however, that a seed of such
a formulation might be found in reference~\Nak.
Lorentzian triangulations  are also interesting in their own right as
a new statistical model, allowing 
for the definition of a new class of lattice
models. Lorentzian triangulations lie somewhere 
between regular lattices with defects
and completely random lattices (i.e.\ dynamical triangulations) 
and it is possible that models based on this
type of lattices will reveal yet other universality classes. 
In this spirit we will study in 
the present paper a generalized model of Lorentzian 
triangulations where the time slice structure is preserved but
where some outgrowths within the time slices are allowed.
\fig{A generalized Lorentzian triangulation in the dual picture. 
}{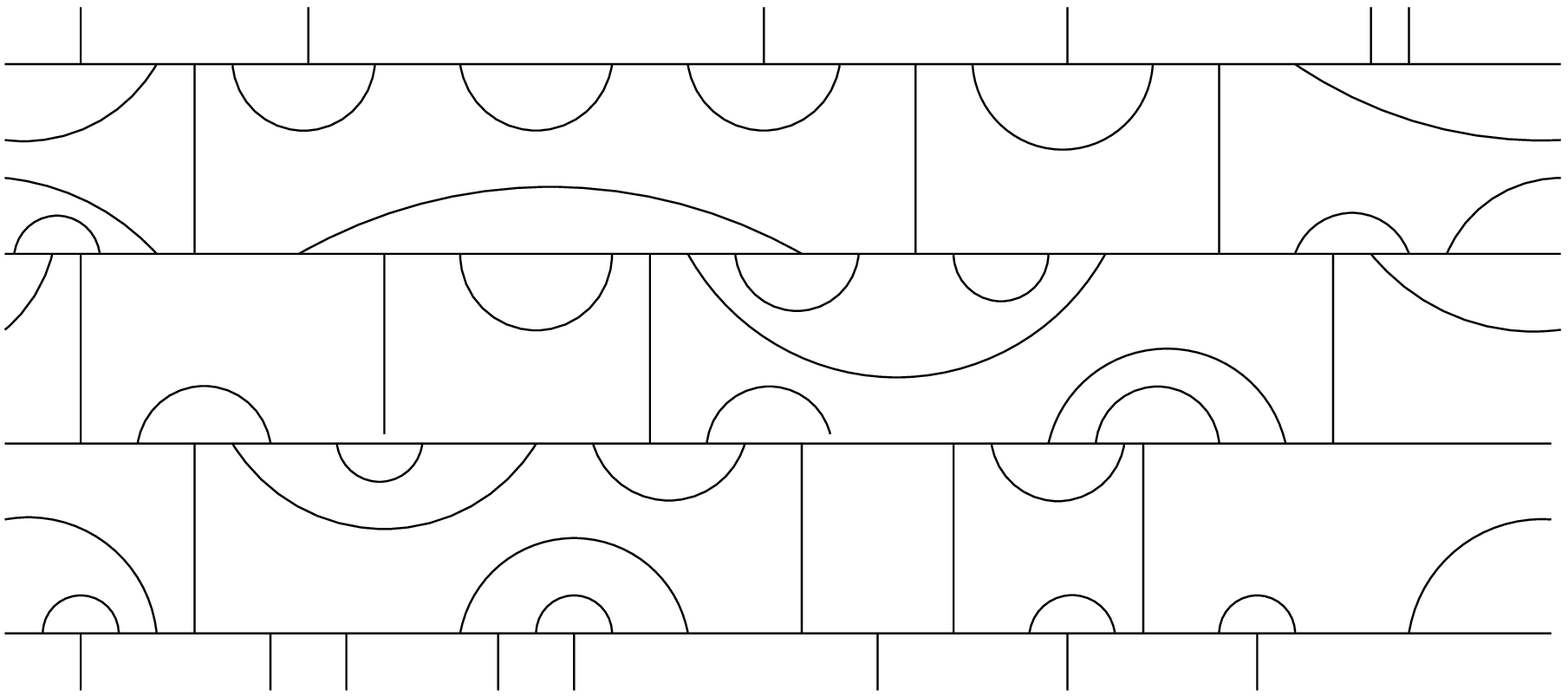}{9.0cm}
\figlabel\archtri
Such a generalized Lorentzian triangulation is displayed in Fig.\archtri\ 
in the dual picture and is obtained by decorating an ordinary Lorentzian 
triangulation such as that of Fig.\lorgra\ (b) by adding arbitrary
space-like ``arches"
connecting two points of the same time line.
These arches do not intersect each other nor the vertical straight lines. 
For later convenience, the boundary condition
imposed on the arch configurations is such that any number of arches
can escape from the right hand side of a given slice and reappear on
the left hand side, see Fig.~\archtri. An elementary arch, in the language
of triangulations, corresponds to a pair of connected
triangles that do not propagate in time, as opposed to pairs of triangles
dual to a vertical edge. 
In the terminology of quantum gravity such an
object constitutes a particular type of (small) ``baby-universe".
Keeping to the statistical mechanical language
we will simply refer to such decorations
as ``outgrowths".
Similarly, more involved arch
configurations correspond to more involved outgrowths each living 
in a single time-slice. 
Clearly, the class of outgrowths we allow for 
constitutes only a small subset of the baby-universes present in the
dynamically triangulated or Euclidean quantum gravity model. In order
to get all baby-universes appearing in Euclidean gravity one 
would have to allow for loops
connected in all possible ways to the already existing elements of the
lattice via three-valent lattices. Such a model has been studied in
ref.~\DK. It corresponds to the fully packed phase of the
so-called $O(1)$ model on a random lattice.
Note finally that the outgrowths we consider
would not suffice to generate surfaces of arbitrary 
topology. The surfaces we consider have genus zero or one depending 
on boundary conditions.
 
Our motivation for studying the particular type of lattices depicted
in Fig.~2 is two-fold. First, in the context of 2D quantum gravity,
it has been shown that one can view Lorentzian quantum gravity as a
renormalized version of Euclidean quantum gravity \ACKL. Obviously, in
going from Euclidean to Lorentzian triangulations baby-universes must
be integrated over.  A key point in the above renormalization
procedure consists in a non-analytical redefinition of the ``boundary
cosmological constant" which is the parameter coupled to the
length of the spatial slices, i.e. the length of  the horizontal lines
in Fig.~1 or Fig.~\archtri. 
In our model all outgrowths lie along the horizontal
lines and one of our aims is to investigate whether the redefinition of the
boundary cosmological constant can be due to these decorations alone.

Secondly, from the statistical mechanical point of view, lattices
of the type depicted in
Fig.~2 are interesting because the system of arch configurations is 
known to be
critical~\cycles~and by defining this system on Lorentzian
triangulations one might be able to change the universality class of
the geometrical system. 
So far one has not been able to study analytically the behavior of
Lorentzian triangulations when critical matter fields are introduced.
One has studied numerically systems consisting of one to eight Ising spins 
on Lorentzian triangulations~\AAL. These studies show that the
interaction between matter and geometry is much weaker than for the models
based on dynamical triangulations. Analytical investigations have been
carried out for Lorentzian triangulations equipped with various dimer fields
and for the case where higher curvature interactions are present~\FGK. These
models all turned out to belong to the same universality class as pure
Lorentzian triangulations. Remarkably, the concept of integrability
generalized to non-regular (Lorentzian) lattices played
a crucial role in their exact solutions and led to the 
hope of extending the standard techniques of integrable lattice models to
models on fluctuating lattices.  

The paper is organized as follows.
In Sect.~2 we describe in more detail the generalized Lorentzian triangulations
that we are going to study and write down their transfer matrix description.
Next, in Sect.~3 we reformulate the model
using a Schwinger-type integral representation leading to a continuous
transfer matrix. 
The Schwinger language provides us with an alternative way of obtaining 
the exact solution in the case of pure Lorentzian triangulations.
In Sect.~4 we consider the case where arches or outgrowths are present 
and show the existence of three different regimes of the model
according to the value of a parameter $\beta$ governing the density of
outgrowths.
In the interesting case of low density of outgrowths ($\beta<1$), 
we define in Sect.~5 a sensible scaling limit in which
our model gets identified with a well-known
quantum mechanical system, namely that governed
by one-dimensional Calogero Hamiltonian. 
Using this equivalence, we compute several interesting thermodynamic
quantities. Finally,
we discuss in Sect.~6 various aspects of our results.

\newsec{The model}

Our generalized model is based on triangular lattices
which in the dual picture look as in Fig.~\archtri. 
In addition to a Boltzmann weight $g$ per triangle,
we introduce a weight $h$ for each triangle which in
the dual language is part of a vertical edge and
a weight $\theta$ for each triangle which in the dual
language is part of an arch, see Fig.~3. 
We also introduce the parameter 
\eqn\betadef{ \beta= {\theta \over h}, }
governing the density of arch decorations.
\fig{The weights associated with the different elements of a
generalized Lorentzian triangulation.}{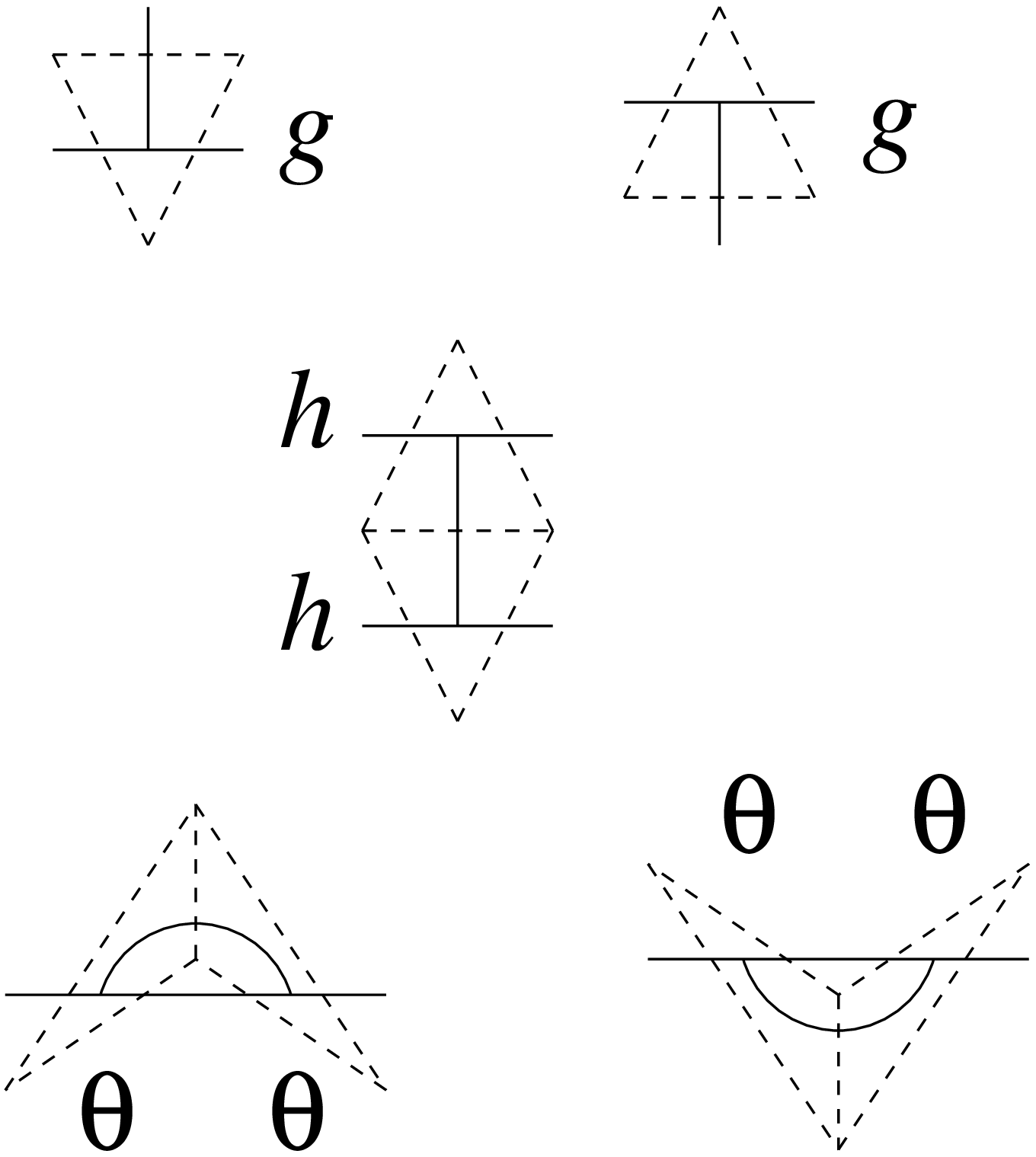}{5.0cm}
\figlabel\couplings
\noindent The partition function of our model over a time lapse $t$ is obtained
by summing over all such generalized triangulations with exactly
$t$ time slices and weighted as just explained.

\fig{The transfer matrix $\Theta$ is decomposed into two parts:
the matrix $\Theta^{(1)}$ of the pure case, and the matrix
$\Theta^{(2)}$ implementing the upper and lower arch 
decorations.}{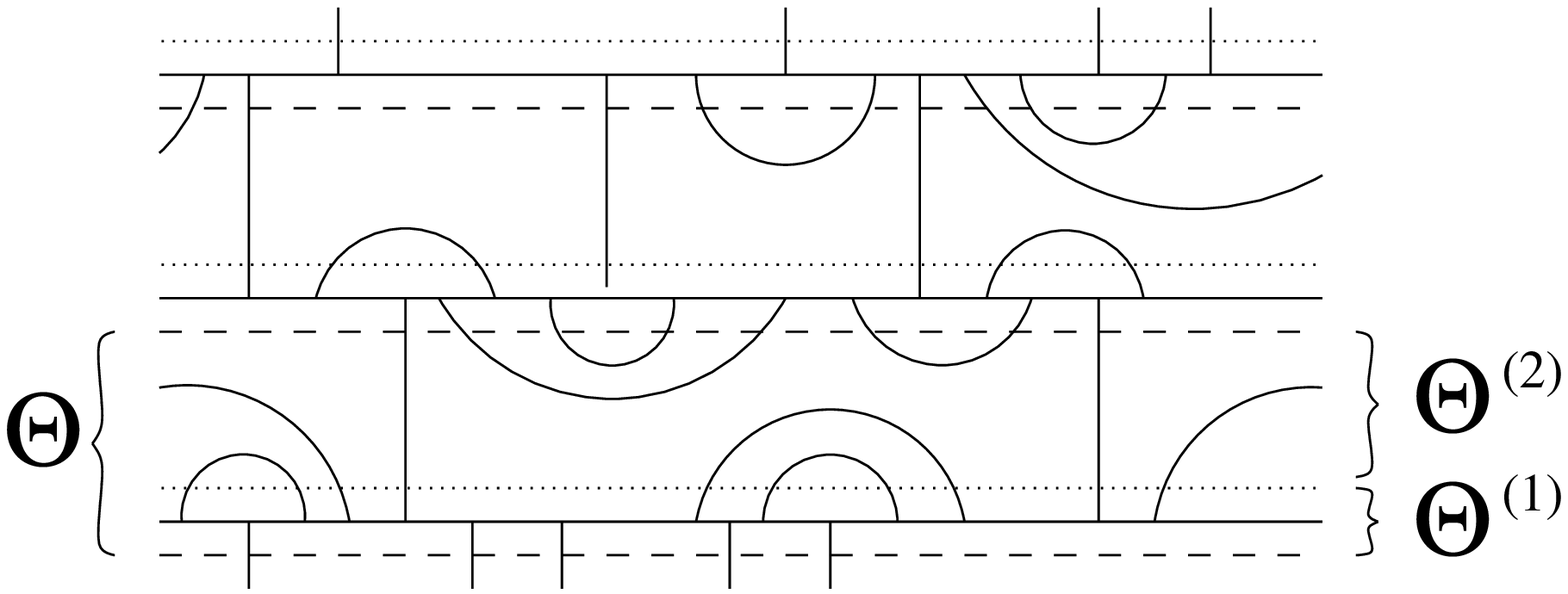}{8.cm}
\figlabel\transmat

As in the case of pure Lorentzian triangulations, 
we may describe our model using a transfer matrix
approach. The transfer matrix describes the effect of adding one
time-slice to our lattice. 
In the present case it is convenient to
view the addition of one slice of the lattice as 
two consecutive steps.
We first apply the transfer matrix of pure Lorentzian triangulations 
$\Theta^{(1)}$
which corresponds to adding the portion of lattice below the dotted line 
and above the dashed line in Fig.~\transmat, including a line of
constant time.   
The transfer matrix element
$\Theta^{(1)}_{ij}$ is indexed by two non-negative integers $i$ and $j$ which
denote the number of half-edges going into the time line and that
coming out of the line respectively and counts the (weighted) arrangements of 
these half-edges along the time line. 
We then complete the slice addition by applying a transfer matrix $\Theta^{(2)}$
which corresponds to adding
the portion of lattice above the dotted line and below
the dashed one in Fig.~\transmat\ and accounts for the pairing of half-edges 
into arches or into vertical edges. The transfer matrix element
$\Theta^{(2)}_{ij}$ is also indexed by the numbers of in and out-coming half
edges.
The total transfer matrix
$\Theta$ can therefore be written as a product of the two above 
transfer matrices
\eqn\composition{\Theta_{ij}=\sum_k\Theta^{(1)}_{ik}\, \Theta^{(2)}_{kj}\ .}
In the case $\theta=0, \ h=1$ of pure Lorentzian triangulations, 
the matrix $\Theta^{(2)}$ reduces to the identity. The new
ingredient is therefore entirely contained in the definition of $\Theta^{(2)}$.

For convenience we will work mainly in the
language of generating functions, by defining
\eqn\Laplace{\Theta(x,y)=\sum_{i,j=0}^{\infty}\Theta_{ij}\, x^i y^j,}
and similarly for $\Theta^{(1)}$ and $\Theta^{(2)}$.
The composition law~\composition~ then reads
\eqn\compLap{\Theta(x,y)=\oint_{\cal C} {d \omega \over 2 \pi i\, \omega}
\Theta^{(1)}(x,{1\over \omega})\Theta^{(2)}( \omega,y),}
with the contour ${\cal C}$ encircling the origin.
The generating functions can be easily derived, with the result
\eqn\thetas{\eqalign{\Theta^{(1)}(x,y)&={1 \over 1-gx-gy}, \cr
\Theta^{(2)}(x,y)&=
{C(\theta^2 x^2) \over 1-\theta^2 x^2 C^2(\theta^2 x^2)}
{C(\theta^2 y^2) \over 1-\theta^2 y^2 C^2(\theta^2 y^2)}
{1 \over 1-h^2 x y C(\theta^2 x^2)C(\theta^2 y^2)},\cr}}
where $C(z)$ is the generating function for Catalan numbers
\eqn\Catalan{C(z)={1-\sqrt{1-4z}\over 2z}\ .}
The result for $\Theta^{(1)}$ is the direct consequence of
the matrix element $\Theta^{(1)}_{ij}$ taking the value 
${i+j\choose i} g^{i+j}$, where we choose for convenience to
include the weight $g$ per triangle in $\Theta^{(1)}$.  
The derivation of $\Theta^{(2)}$ is slightly more involved: we
attach to each vertical edge the upper and lower arch configurations
sitting immediately to its right (see Fig.\transmat). 
This results in an effective weight
$h^2 x y C(\theta^2 x^2)C(\theta^2 y^2)$ per vertical edge. Indeed,
$C(z)$ is the generating function for arbitrary arch configurations
with a weight $z$ per arch. Summing over the number of vertical edges
leads to the third factor in the second line of \thetas.
The first two factors come from our particular choice of boundary conditions.
Each is obtained by summing over the number of escaping arches each effectively
weighted by a factor $\theta^2 z^2 C^2(\theta^2 z^2)$, with $z=x$ resp.\ $y$
for upper, resp.\ lower escaping arches. We remark that as opposed to
what is the case for $\Theta^{(1)}$ and $\Theta^{(2)}$, the transfer
matrix $\Theta$ is not symmetric with respect to $x$ and $y$.\foot{
This asymmetry of $\Theta(x,y)$
can be cured by, instead of
defining $\Theta$ by~\composition, setting
$\Theta=(\Theta^{(1)})^{1/2} \Theta^{(2)} (\Theta^{(1)})^{1/2}$. 
Since the transfer matrix of the pure model has
been explicitly diagonalized~\FGK~it is straightforward to write down an
expression for $(\Theta^{(1)})^{1/2}$. However, the full transfer
matrix $\Theta_{ij}$ resulting from this definition has no longer 
a clear geometrical interpretation as being associated
with $i$ incoming edges and $j$ outgoing ones. 
In appendix A we show how to calculate 
$\Theta=(\Theta^{(1)})^{1/2} \Theta^{(2)} (\Theta^{(1)})^{1/2}$. Here
we shall take a line of action which does not require the knowledge
of the exact transfer matrix.}

Among other quantities of interest, we will compute the
$t$-step partition function with periodic boundary conditions in the
time direction
\eqn\Zt{Z(t)=\Tr (\Theta^t).}
We will also calculate correlation functions for the total number
${N}(s)$ of triangles in the slice at time $s$, with the same $t$-step
periodic boundary conditions.  

Apart from these we will be interested in
calculating the $t$-step partition function with {\it open} boundaries
in the $t$-direction $Z(i,j,t)$
\eqn\Zij{Z(i,j,t)=\left((\Theta^{(1)}\Theta^{(2)})^t\,\Theta^{(1)}\right)_{ij},}
or rather its associated generating function $Z(x,y,t)$.
With a slight abuse of language we will also refer to $Z(x,y,t)$ or $Z(i,j,t)$
as the loop-loop correlator.
Notice that $Z(i,j,t)$
is not simply the $t$-th power of the transfer matrix. 
It is, however, to our opinion the most natural
definition of a $t$-step partition function with open boundaries in
the $t$-direction. It describes
triangulations for which the boundary conditions in the $t$-direction
are as depicted in Fig.~\archtri. There are $i$ incoming vertical edges and
$j$ outgoing ones and no arches are allowed either between incoming or
outgoing edges.

For pure Lorentzian triangulations, two different strategies were used for  
calculating objects as given above. One strategy consists in first determining
$\Theta^t(x,y)$ by writing down and solving a recursion relation in
$t$ for this quantity~\AL. 
The other one consists in explicitly
diagonalizing the transfer matrix $\Theta_{ij}$~\FGK. 
Neither
method appears to be tractable for the present problem 
with general $\theta$ and $h$.  
We will resort to yet another approach using an integral representation
of the transfer matrix.

\newsec{Schwinger representation}

\subsec{Transfer matrix}

In this section we will reformulate the discrete
transfer matrix $\Theta$ with indices $i,j=0,1,2,...$
as a continuous transfer kernel by means of an integral representation, using
continuous Schwinger parameters $\alpha, \alpha' \in [0,\infty)$. 
This representation is inspired by the calculation in ref. \DK\ of the
``watermelon" correlation function of the $O(n)$ model coupled
to 2D Euclidean gravity.
For definiteness, let us consider the partition function
$Z(t)$ \Zt. From~\compLap~it follows that 
\eqn\defpart{Z(t)=
\prod_{s=1}^t\bigg(\oint_{\cal C}{d w_s \over 2i\pi w_s}
{d z_s \over 2i\pi z_s}\bigg) \prod_{r=1}^t 
\Theta^{(1)}\left({1\over w_{r-1}},{1\over z_r}\right)\;
\Theta^{(2)}(z_r,w_r),}
where the contour is a small circle around the origin and where we
have imposed cyclic boundary conditions on the $w$'s, namely  
$w_{0}=w_t$.

Whereas in the pure case the integrand in~\defpart~would have
singularities only in the form of poles, in the present case the
integrand in addition has singularities in the form of cuts. We can,
however, transform the cuts into poles by the following change of
variables
\eqn\chvar{
\mu_r=\theta w_r C(\theta^2 w_r^2), \qquad
\rho_r=\theta z_r C(\theta^2 z_r^2).}
This change of variables maps the circle ${\cal C}$ into a closed
ellipse-like curve
${\cal E}$ still encircling the origin.  Using the quadratic equation
satisfied by the Catalan generating function
\eqn\quadratic{x (C(x))^2=C(x)-1,}
the relation~\chvar~ is easily  inverted into 
\eqn\invch{ w_r={1\over \theta (\mu_r+{1 \over \mu_r})}, \qquad
z_r={1\over \theta (\rho_r+{1 \over \rho_r})}.}
{}From there it follows that
\eqn\measure{{dw_r \over w_r^2}={\theta (1-\mu_r^2)d\mu_r \over \mu_r^2},}
and a similar relation holds between $dz_r$ and
$d\rho_r$. Inserting~\chvar~and~\measure~in~\defpart~we get
\eqn\newpart{Z(t)=\prod_{s=1}^t\bigg(\oint_{\cal E}
{d \mu_s \over 2 i \pi \mu_s}{d \rho_s \over 2 i \pi \rho_s}\bigg)
\prod_{r=1}^t {1 \over 1-{\rho_r\mu_r\over \beta^2}} 
{1\over 1-g\theta(\rho_r+{1\over \rho_r}+\mu_{r-1}+{1\over
\mu_{r-1}})},}
where $\beta=\theta /h$ as in \betadef.
We may next introduce a Schwinger representation for the second factor
in each of the products in the integrand. This gives
\eqn\schwin{ Z(t)=\prod_{s=1}^t\bigg(\oint_{\cal E}
{d \mu_s \over 2 i \pi \mu_s}{d \rho_s \over 2 i \pi \rho_s}\int_0^\infty 
d\alpha_s e^{-\alpha_s}\bigg)
\prod_{r=1}^t {1 \over 1-{\rho_r\mu_r\over \beta^2}}e^{g\theta\alpha_r
(\rho_r+{1\over \rho_r}+\mu_{r-1}+{1\over \mu_{r-1}})}. }
The integrals over the $\rho_r$ may now be performed, by noticing that
each of them simply picks the residue of 
the integrand at $\rho_r=0$
This gives
\eqn\muint{ Z(t)=\prod_{s=1}^t\bigg(\oint_{\cal E}
{d \mu_s \over 2 i \pi \mu_s}\int_0^\infty 
d\alpha_s e^{-\alpha_s}\bigg)
\prod_{r=1}^t \sum_{k\geq 0}I_k(2g\theta\alpha_r)
\left({\mu_r \over \beta^2}\right)^ke^{g\theta\alpha_{r+1}
(\mu_{r}+{1\over \mu_{r}})}, }
where the $I_k$'s are the modified Bessel functions of the first kind, i.e.\
$I_k(2 x)=\sum_{p\geq 0} x^{2p+k}/(p!(p+k)!)$ and
where we have imposed periodic boundary conditions on the $\alpha$'s, i.e.\
$\alpha_{t+1}=\alpha_1$.

Similarly, to carry out the 
integral over $\mu_r$ we simply have to pick the residue of the
integrand at $\mu_r=0$ and the expression finally reduces to
\eqn\lutfin{Z(t)=
\prod_{s=1}^t\bigg(\int_0^\infty d\alpha_s e^{-\alpha_s}\bigg)
\prod_{r=1}^t \phi_\beta(g\theta\alpha_{r},g\theta\alpha_{r+1}), }
where we have defined the transition function
\eqn\transit{ \phi_\beta(x,y)=\sum_{k\geq 0} I_k(2 x)
I_k(2 y)/\beta^{2k}.}
Note that as expected the partition function depends on
$g$, $h$ and $\theta$ only via $g\theta$ and $\beta=\theta / h$.

Eqn. \lutfin\ leads to the definition of a continuous symmetric
transfer kernel 
\eqn\noyau{ G_{\beta,g\theta}(\alpha,\alpha')= 
e^{-{\alpha+\alpha'\over 2}} \ \phi_\beta(g \theta \alpha, g \theta \alpha'),}
to be integrated
with the flat measure on the parameters $\alpha \geq 0$.
To make contact with the original discrete formulation of the problem,
let us show how to obtain statistical properties involving the observable
${N}(s)=i+j$ counting the total number of triangles in the slice 
at time $s$ ($i$ and $j$ denote respectively the number of in- and out-coming
half edges at the time-line $s$), from the 
statistical properties of the variable $\alpha_s$.  
Let us define a more general local observable $\Sigma_s(z)$ through 
\eqn\factomo{\eqalign{
\sigma_n(s)&= {N}(s) ({N}(s)-1)...({N}(s)-n+1),\cr
\Sigma_s(z)&= \sum_{n=0}^\infty \sigma_n(s) {z^n \over n!}\ . \cr}}
The correlations of $\Sigma$ at various times $s_1,...,s_k$ read simply
\eqn\corsig{ \langle \prod_{m=1}^k \Sigma_{s_m}(z_m) \rangle_{\Theta} =
\langle \prod_{m=1}^k {e^{{\alpha_{s_m} z_m\over 1+z_m}} \over 1+z_m} 
\rangle_G,}
where the subscripts $\Theta,\ G$ refer to the framework (discrete, continuum)
in which the correlation is evaluated. This is readily proved by noticing that
the desired correlator corresponds to substitutions $g \to g (1+z_m)$ in
the formula for the partition function \lutfin\ within the slice $s_m$. 
The latter can then be absorbed into a change of variables
$\alpha_{s_m} \to \alpha_{s_m}/(1+z_m)$.
This will be extensively used in Sect.~5 below.

\subsec{Example 1: pure Lorentzian triangulations revisited}

As a preliminary exercise, it is instructive to re-derive the exact results
of pure Lorentzian triangulations in this new continuum language.
Let us diagonalize the continuum transfer matrix of pure  
Lorentzian triangulations (with $\theta=0$ and $h=1$)
\eqn\defGzero{ G^{(0)}(\alpha,\alpha')=\lim_{\beta \to 0} G_{\beta,
g \beta}(\alpha,\alpha') =
e^{-{\alpha+\alpha'\over 2}}\ I_0( 2 g \sqrt{\alpha \alpha'}) ,} 
where we have used that $I_k(2g \beta x)/\beta^k \to (gx)^k/k!$ in \transit.
It turns out that the eigenfunctions of $G^{(0)}$ are simply 
related to the Laguerre polynomials $L_n(x)$, orthogonal wrt.\ the measure
$e^{-x} dx$ over the positive reals. These polynomials read
\eqn\lag{ L_n(x)= \sum_{m=0}^n {n \choose m} {(-x)^m \over m!}. }
The diagonalization follows from the following quadratic formula
\eqn\meixL{ \sum_{n=0}^\infty e^{-{x\over 2}}L_n(x) e^{-{y \over 2}}L_n(y)
u^{2n+1} = {u \over 1-u^2} e^{-{1\over 2}{1+u^2\over 1-u^2}(x+y)} 
I_0\left(2 {u \sqrt{xy}\over 1-u^2}\right) ,}
valid for $u<1$.
Hence upon setting $g=1/(q+1/q)$,  
and defining $\alpha=x(1+u^2)/(1-u^2)$,
$\alpha'=y(1+u^2)/(1-u^2)$, with $u=q$, eqn. \meixL\ translates into 
\eqn\digGzero{ G^{(0)}(\alpha,\alpha')= 
\sum_{n=0}^\infty \psi_n^{(0)}(\alpha)\psi_n^{(0)}(\alpha')
\lambda_n ,}
in which $\psi_n^{(0)}(\alpha)$ is the normalized 
eigenfunction for the eigenvalue
$\lambda_n$, respectively reading
\eqn\psinordef{ \eqalign{
\psi_n^{(0)}(\alpha)&= \sqrt{1-q^2 \over 1+q^2} e^{-{1\over 2}{1-q^2\over 
1+q^2}\alpha}
L_n\left({1-q^2\over 1+q^2}\alpha\right), \cr
\lambda_n&= {q^{2 n+1} \over g} , \cr
q&= g C(g^2) , \cr}}
with $C$ as in \Catalan. 

This leads immediately to the pure Lorentzian triangulation partition function
on a time cylinder with $t$ steps 
\eqn\Zzero{ Z^{(0)}(t)= \sum_{n=0}^\infty \lambda_n^t = {1\over g^t} {q^t \over
1-q^{2t}}= {C(g^2)^t \over 1-[g C(g^2)]^{2t}}. }
Similarly, we get the one-point correlation of the observable
$\Sigma_s(z)$ as in \corsig\ by using the formula \meixL\ with
$x=y=\alpha (1-q^2)/(1+q^2)$ and $u=q^t$:
\eqn\corsizer{\langle \Sigma_s(z)\rangle_\Theta= 
{1\over 1+z}\langle e^{{\alpha z \over 1+z}} \rangle_{G^{(0)}}
={ I(z)\over I(0)}, }
where 
\eqn\defI{ I(z)={1\over 1+z} \int_0^\infty d \alpha e^{\alpha\big( 
{z \over 1+z} -
{1+q^{2t}\over 1-q^{2t}}
{1-q^2\over 1+q^2} \big) } 
I_0\left(2\alpha {q^t \over 1-q^{2t}}
{1-q^2\over 1+q^2} \right) .} 
Performing explicitly the integral, we find
\eqn\fincorzer{
\langle \Sigma_s(z)\rangle_\Theta= {1 \over \sqrt{1+z- z{1+q^t\over
1-q^t}{1+q^2\over 1-q^2}}} {1\over \sqrt{1+z- z{1-q^t\over
1+q^t}{1+q^2\over 1-q^2}}} .}

The expressions \Zzero\ and \fincorzer\ have a singularity at 
$q=1$, i.e. $g=1/2$. In order to obtain a
continuum theory, we therefore write
\eqn\scalpure{2g=1-{1\over 2}a^2\Lambda,}
where $a^2$ is a scaling parameter with the dimension of area and
where $\Lambda$ is the renormalized fugacity per triangle. Using this
scaling for $g$ we have
\eqn\scalC{q=g\, C(g^2)\sim e^{-a\sqrt{\Lambda}},}
and we see that in order to obtain a finite expression for $Z^{(0)}(t)$ in
the continuum limit we must set
\eqn\scaltpure{t={T\over a}.}
This gives the following continuum partition function $Z_T^{(0)}$
\eqn\Zcont{Z_T^{(0)}\equiv \lim_{a\to 0}{1\over 2^t} Z^{(0)}(t)
={e^{-\sqrt{\Lambda}T}\over
1-e^{-2\sqrt{\Lambda}T}}.}
Similarly, we must take 
\eqn\scalz{ z= a \sqrt{\Lambda} {\cal Z} \ \ \ {\rm and}\ \ \ N={{\cal N}\over 
a}, }
where the factor $\sqrt{\Lambda}$ is simply for convenience (${\cal Z}$
is dimensionless), and substitute this into
\fincorzer\ to get the scaled average 
\eqn\scacozer{ \langle e^{ \sqrt{\Lambda}{\cal N}{\cal Z} } \rangle =
{1\over \sqrt{1-2 {\cal Z}{\rm cotanh}(\sqrt{\Lambda} T)+{\cal Z}^2 }}, }
generating the moments $\langle {\cal N}^k\rangle$
of the (rescaled) number of triangles per time slice.
These moments read
\eqn\legendre{ \langle {\cal N}^k\rangle = 
{k!\over \Lambda^{k\over 2}} P_k\big({\rm cotanh}(\sqrt{\Lambda}T)\big)
={k!\over \Lambda^{k\over 2}}
{\sum_{m=0}^k {k \choose m}^2 e^{-2m \sqrt{\Lambda}T}
\over (1-e^{-2 \sqrt{\Lambda}T})^k } ,}
where $P_k(x)$ denotes the $k$-th Legendre polynomial.

\subsec{Example 2: the case $\beta=\infty$  } 

For $\beta \to \infty$ the continuum transfer kernel
factorizes as follows
\eqn\betinf{ G_{\infty,g \theta}(\alpha,\alpha')=
e^{-{\alpha\over 2}} I_0(2 g \theta \alpha)\times e^{-{\alpha'\over 2}} 
I_0(2 g \theta \alpha'). }
The partition function therefore factorizes too and reads
\eqn\zlinf{\eqalign{ Z^{(\infty)}(t)&= (Z^{(\infty)}(1))^t, \cr
Z^{(\infty)}(1)&=\int_0^\infty d\alpha e^{-\alpha} I_0(2 g \theta \alpha)^2
={2\over \pi}K((4g\theta)^2), \cr}}
where $K(x)$ is the complete elliptic integral of the first kind. 
Note that $Z^{(\infty)}(1)$ is simply the partition function of two 
(upper and lower) interlocking arch
systems connecting points by pairs along a single time-line.  
The function $K(x)$ is singular at $x=1$. In order to define a continuum
theory we would therefore in this case set
\eqn\scalinf{4g\theta=1-a^2 \Lambda.}
Then we get
\eqn\Zone{Z^{(\infty)}(1)={2\over \pi}K((1-a^2\Lambda)^2)=-{1\over\pi}{\rm Log}
(a^2\Lambda) +{\cal O}(1),}
and the leading singular behavior of $Z^{(\infty)}(t)$ is therefore
\eqn\zinfsing{Z^{(\infty)}(t)\sim \left(-{1\over\pi}{\rm 
Log}(a^2\Lambda)\right)^t.}
Thus, in this case it is {\it not} possible to define a sensible continuum
time variable.

\newsec{Scaling regimes}

It is obvious that the possible singularities of $Z(t)$ must come from
large Schwinger parameters, $\alpha_r$, since the Bessel functions
$I_k(2x)$ are polynomial for small $x$ and behave for large $x$ as
\eqn\behambes{ I_k(2 x) \simeq {e^{2 x} \over \sqrt{4 \pi x}}
\left(1-{4k^2-1\over 16 x}+O\left({1\over x^2}\right)\right).}
We will therefore be interested in determining the asymptotic
behavior of the transfer kernel $G_{\beta,g\theta}(\alpha,\alpha')$ as 
$\alpha$, $\alpha'$ $\to \infty$. 
It turns out that one has three different scaling
regimes corresponding respectively to $\beta<1$, $\beta=1$ and $\beta>1$.

\subsec{Three scaling regimes}

When $\beta>1$, the transition function $\phi_\beta(x,y)$ in
\noyau\ is expressed as 
an absolutely convergent series \transit, that behaves for large $x$
and $y$ as
\eqn\phibeha{ \phi_\beta(x,y) \sim {e^{2(x+y)}\over 4 \pi \sqrt{xy}}
\sum_{k\geq 0}{1\over \beta^{2k}} ={\beta^2\over \beta^2-1}
{e^{2(x+y)}\over 4 \pi \sqrt{xy}}. }
In order to determine the asymptotic behavior of
$\phi_{\beta}(x,y)$ for the remaining values of $\beta$ we make use
of the generating function for modified Bessel functions of the first kind
\eqn\mobes{ e^{x(t+{1\over t})}=I_0(2 x)+\sum_{k\geq 1}
(t^k+{1\over t^k})I_k(2 x).}
{}From this relation it follows that
\eqn\invphi{\eqalign{ \oint {dt\over 2 i \pi t}e^{x(t+{1\over t})
+y(\beta^2t+{1\over \beta^2t})} &=\phi_\beta(x,y) +\phi_{1\over \beta}(x,y)
-I_0(2x)I_0(2y)\cr
&=I_0\left(2 \sqrt{(x+\beta^2 y)(x+y/\beta^2)}\right).\cr}}
For $\beta=1$ we immediately get from~\invphi\
\eqn\comphi{ \phi_1(x,y)= {1\over 2}\big(I_0(2(x+y))+I_0(2 x)I_0(2 y)\big),}
which behaves for large $x$ and $y$ as 
\eqn\beone{ \phi_1(x,y)\sim {e^{2(x+y)} \over 4 \sqrt{\pi(x+y)}}. }
Finally, when $\beta<1$, we write \invphi\ as
\eqn\crypt{ \phi_\beta(x,y)=I_0\left(2 \sqrt{(x+\beta^2
y)(x+y/\beta^2)}\right)+I_0(2x)I_0(2y)-\phi_{1\over \beta}(x,y),} 
and notice that since
$\phi_{1\over \beta}(x,y)$ is now the absolutely convergent
series, the last two terms in \crypt\ behave like $e^{2(x+y)}$.
Using moreover that 
\eqn\domin{\sqrt{(x+\beta^2 y)(x+y/\beta^2)}>x+y,}
for all real $\beta\neq 1$,
the large $x,y$ asymptotics of $\phi_\beta(x,y)$ is entirely governed by 
the first term on the rhs.\ of \crypt.
This means that for $\beta<1$ we have the following asymptotic
behavior of the transition function as $x,y$ $\to \infty$
\eqn\behainv{ \phi_\beta(x,y)\sim 
{e^{2 \sqrt{(x+\beta^2 y)(x+y/\beta^2)}}\over 
2\sqrt{\pi \sqrt{(x+\beta^2 y)(x+y/\beta^2)}}} \left( 1 +
{1\over 16 \sqrt{(x+\beta^2 y)(x+y/\beta^2)}} +...
\right). } 

To summarize, we have found the following dominant behavior for 
the transfer kernel $G_{\beta,g\theta}(\alpha,\alpha')
\sim e^{-S_{\beta,g\theta}(\alpha,\alpha')}$ as $\alpha,$ $\alpha'$
$\to \infty$. 
\eqn\tkG{ \eqalign{
\beta \geq 1 \ : \ S_{\beta,g\theta}(\alpha,\alpha')&={1\over 2} 
(\alpha+\alpha')(1-4 g\theta), \cr
\beta < 1 \ : \ S_{\beta,g\theta}(\alpha,\alpha')&={1\over 2} 
(\alpha+\alpha') -2 g\theta\sqrt{(\alpha+{\alpha'\over \beta^2})
(\alpha+\beta^2\alpha')}, \cr}}
where the cases $\beta>1$ and $\beta=1$ differ by sub-leading corrections. 
A crucial difference between $\beta<1$ and $\beta\geq 1$ is that in the latter
case the partition function
is factorized (at leading order), while in the former case it will be dominated
by correlated $\alpha$'s all of the same order.
This in turn reflects the natural property that 
for low arch densities, successive time slices have lengths of the same order,
while the presence of many arches allows for decorrelated lengths.

The following three sections are devoted to the 
three cases $\beta=1$, $\beta>1$
and $\beta<1$ for finite $t$. 
In all cases, we will need the large-$\alpha$ asymptotics
of the product of $t$ transfer kernels which may be expressed as
$e^{-S_\beta}\times U_\beta$, with an action 
$S_{\beta}(\{\alpha_s\})$, reading
\eqn\actnf{S_\beta(\{\alpha_s\})=\sum_{s=1}^t 
S_{\beta,g \theta}(\alpha_{s},\alpha_{s+1}) , }
where $\alpha_{t+1}\equiv \alpha_1$, 
and a non-exponential factor $U_\beta(\{\alpha_s\})$, which will be detailed
below.

\subsec{The case $\beta=1$}

The action $S_1(\{\alpha_s\})$ and the pre-factor $U_1(\{\alpha_s\})$
read respectively
\eqn\actone{S_1(\{\alpha_s\})=(1-4 g\theta)\sum_{i=1}^t \alpha_s, }
and
\eqn\sublone{ U_1(\{\alpha_s\})={1\over (16\pi g \theta)^{t\over 2}}
\prod_{s=1}^t {1\over \sqrt{\alpha_s+\alpha_{s+1}}}. }
{}From the action \actone, we immediately find the singularity of $Z(t)$
to be at 
\eqn\singone{4g \theta =1.}
Furthermore, writing
$4 g \theta=1- a^2 \Lambda$ and performing the change
of variables $\alpha_s=\beta_s/a^2$ in the integral \lutfin\
with $\beta=1$, we get
\eqn\singbehaone{ Z(t) \sim
{1 \over a^{t}} \int_0^\infty \prod_{s=1}^t {d\beta_s
e^{-\Lambda \beta_s} \over 2\sqrt{\pi}\sqrt{\beta_s+\beta_{s+1}}}, }
when $a \to 0$. Note that the multiple integral converges and
gives a non-trivial dependence on $t$.

This calculation should be compared to that of \DK\ computing the
singularity structure of the $t$-slice
``watermelon" correlator of the
$O(n)$ model coupled to ordinary 2D Euclidean gravity. 
Indeed,
viewing the slices of watermelon as time-slice separators, the correlator
becomes in the limit $n\to 0$ equivalent to a periodic time Lorentzian
partition function with $h=\theta$
and with slightly different boundary conditions (no escaping
arches). 
In both cases the singularity of the partition function is factorized,
leading in the case of ref. \DK\ to a different behavior $Z\sim a^t$, in
agreement with the KPZ scaling \KPZ.

\subsec{The case $\beta>1$}

In this case, the action is the same as in the $\beta=1$ case
$S_\beta=S_1$ \actone, while the pre-factor $U_\beta(\{\alpha_s\})$
reads
\eqn\subltwo{ U_\beta(\{\alpha_s\})=
\left(\beta^2\over (\beta^2-1) 4\pi g \theta\right)^t
\prod_{s=1}^t {1\over \alpha_s}. }

The critical point is therefore the same as in the case $\beta=1$,
namely 
\eqn\singb{4g\theta=1,}
and setting again $4 g \theta=1- a^2 \Lambda$,
the singularity of $Z(t)$ is factorized with the result
\eqn\factotwo{ Z(t) \sim \left( -{1\over \pi} 
{\beta^2\over \beta^2-1}\  {\rm Log} (a^2 \Lambda) \right)^t,}
which, up to a proportionality factor, is identical to the 
behavior at $\beta=\infty$ \zinfsing.

We may infer that this behavior is characteristic of the
$\beta>1$ regime, and
breaks down at $\beta=1$ as \factotwo\ indicates. Once again, for
$\beta>1$, it is not possible to introduce a sensible continuum time
variable.

\subsec{The case $\beta<1$}

In this last case, the dominant action reads
\eqn\actinf{S_\beta(\{\alpha_s\})= \sum_{s=1}^{t} \alpha_s 
-2 g \theta\sum_{s=1}^{t}\sqrt{(\alpha_s+\beta^2\alpha_{s+1})
(\alpha_s+{\alpha_{s+1}\over \beta^2})}, }
and the sub-leading factor is
\eqn\sublead{\eqalign{ U_\beta(\{\alpha_s\})=& 
{1 \over (4 \pi g \theta)^{t\over 2}}\prod_{s=1}^{t} {1\over \big(
(\alpha_s+\beta^2\alpha_{s+1})(\alpha_s+{\alpha_{s+1}\over 
\beta^2})\big)^{1\over 4}} \ \ \times \cr
&\left(1+{1 \over 16 g \theta
\sqrt{(\alpha_s+\beta^2\alpha_{s+1})(\alpha_s+{\alpha_{s+1}\over
\beta^2})}}+...\right),\cr}  }
where in both expressions we assume periodic boundary conditions
for the $\alpha$'s, i.e.\ $\alpha_{t+1}=\alpha_1$.
Let us now turn to studying the singular behavior of $Z(t)$. 
To obtain the critical point, we notice
that the dominant action \actinf\ is homogeneous of degree one in the 
$\alpha$'s, 
thus in order for the partition function to be defined, 
all configurations $\{\alpha \}$ must 
satisfy $S_{\beta}(\{\alpha\})\geq 0$. 
The critical point corresponds to the existence of configurations
$\{\alpha_* \}$ minimizing the action, while
satisfying $S_{\beta}(\{\alpha_*\})=0$. 
These can be obtained from  the saddle-point equations
\eqn\sysmin{
1-{g \theta\over \beta}\left( \beta^2 x_s+{1\over x_s} +x_{s-1}+{\beta^2
\over x_{s-1}}\right)=0, \ \ \ \ \ \ \ \  s=1,2,\ldots,t,} 
where
\eqn\xs{x_s=\left({\alpha_s+\beta^2\alpha_{s+1}\over 
\beta^2\alpha_s+\alpha_{s+1}}\right)^{1/2}
, \ \ \ \ \  x_{s+t}\equiv x_s\ \ {\rm and} \ \ 
\alpha_{s+t}=\alpha_s.}
The only periodic solution $\alpha_s$ to the above corresponds
to $x_s=$ const. $=1$ for all $s$, implying that all the $\alpha$'s 
must be equal, and moreover this  
forces $g \theta$ to take the critical value
\eqn\critipt{ 2 g \theta \left(\beta+{1\over \beta}\right) =1. }
As expected the action at the saddle-point
vanishes automatically.

To capture the singularity of $Z(t)$, let us explore the vicinity of the
critical point \critipt. In analogy with the pure
triangulation case, we set
\eqn\betascal{2 g\theta\left(\beta+{1 \over \beta}\right)=
1-{1 \over 2}a^2 \Lambda,}
where $a^2$ is a small parameter with the dimension of area.
The homogeneous nature of the saddle-point
above, where all the $\alpha$'s are equal, suggests that we 
keep, say $\alpha_1$, unchanged, and that we 
perform the change of variables $\alpha_s=\alpha_1+\xi_s\sqrt{\alpha_1}$
for $s=2,3,...,t$.
Then the sub-leading factor \sublead\ together with the integration
measure behave for large $\alpha_1$ as
\eqn\leadpref{ \prod_{s=1}^t d \alpha_s U_\beta(\{\alpha_s\})\simeq 
{d\alpha_1\over \sqrt{2\pi \alpha_1}} \prod_{s=2}^t 
{{d\xi_s \over \sqrt{2 \pi }}},}
while the action reads
\eqn\actxi{\eqalign{
S(\alpha_1,\{\xi\})&= t\alpha_1+ \sqrt{\alpha_1}
\sum_{s=1}^t \xi_s \cr
&-(1-{1 \over 2} a^2 \Lambda )\alpha_1 \sum_{s=1}^t \sqrt{ 
\big(1+{1 \over \sqrt{\alpha_1}}{\xi_s+\beta^2\xi_{s+1}
\over 1+\beta^2}\big)\big(1+{1 \over \sqrt{ \alpha_1}}
{\xi_{s+1}+\beta^2\xi_s\over 1+\beta^2}\big)} \cr
&={a^2 \over 2} \Lambda t\alpha_1
+{1\over 8}\left({1-\beta^2\over 1+\beta^2}\right)^2 
\sum_{s=1}^t(\xi_{s+1}-\xi_s)^2+O\left({1\over \sqrt{\alpha_1}}\right),\cr} }
where we have set $\xi_1=\xi_{t+1}=0$.
To evaluate $Z(t)$, we must first integrate over the $\xi$'s (along the 
real line)
and then over $\alpha_1>0$. This results in
\eqn\reszlop{\eqalign{
Z(t)&\sim \int_0^\infty {d\alpha_1\over \sqrt{2\pi \alpha_1}}
e^{-{1\over 2}a^2 \Lambda t \alpha_1} 
\left( 2{1+\beta^2\over 1-\beta^2}\right)^{t-1} t^{-1/2} \cr
&={1-\beta^2\over 2 a \sqrt{\Lambda} t (1+\beta^2)} 
\left( 2{1+\beta^2\over 1-\beta^2}\right)^t,\cr}}
where the $\xi$ integral has produced the term $t^{-1/2}=\det(\Delta)^{-1/2}$,
where $\Delta$ is the $(t-1)\times(t-1)$ matrix of the discrete Laplacian,
namely $\Delta_{r,s}=2 \delta_{r,s} -\delta_{r,s+1}-\delta_{r,s-1}$, truncated
to $1\leq r,s \leq t-1$.
Note that in \reszlop\ we have displayed the trivial entropic factor
$(2 (1+\beta^2)/(1-\beta^2))^t$, namely a weight $2$ for each application
of $\Theta^{(1)}$ and $(1+\beta^2)/(1-\beta^2)$ for each application of
$\Theta^{(2)}$. These factors could
of course be absorbed
in a redefinition $\Theta^{(1)} \to \Theta^{(1)}/2$ and
$\Theta^{(2)}\to \Theta^{(2)}/\left({1+\beta^2 \over 1-\beta^2}\right)$.

\newsec{The scaling limit and the Calogero Hamiltonian}

In order to obtain a finite result when $a\to 0$ for the partition function
\reszlop\ (after stripping it from the trivial entropic factor),
we must let simultaneously $t$ scale as $1/a$. We therefore
introduce a continuum time variable  $T$ through
\eqn\Tcontbeta{t={T\over a}.}
We note that~\Tcontbeta\ is the same scaling of the time variable as
in the pure case (cf.\ eqn.~\scaltpure). This means in particular
that the fractal dimension, $d_H$ of our surfaces for
$\beta<1$ is also the same as in the pure Lorentzian 
case, i.e.\ $d_H=2$.

\subsec{Partition function}

Let us now
compute the partition function of the $\beta<1$ model
in the vicinity of the critical point using \betascal\ and \Tcontbeta\
and letting $a \to 0$.
We know already that the integral \lutfin\ should be dominated in this 
limit by large $\alpha$'s that are close to one another. This suggests
that we
directly perform the following
change of variables in the integral 
\eqn\chvaint{ \alpha_s = {\varphi_s^2\over a}, \ \ \ \ \
\ \ \varphi_s>0.}
Inserting the change of variables in the action \actinf\ leads to
\eqn\actnew{\eqalign{ 
S(\varphi)&= {1\over a}\sum_{s=1}^t \varphi_s^2 -{\left(1-{1\over 2}
a^2\Lambda\right) \over
a} \sum_{s=1}^t \varphi_s^2 \sqrt{\big( 1 +
{\varphi_{s+1}^2-\varphi_s^2\over \varphi_s^2 (1+\beta^2)} \big)\big( 1 +
\beta^2{\varphi_{s+1}^2-\varphi_s^2\over \varphi_s^2 (1+\beta^2)} \big)}\cr
&\simeq{1\over 2} a \Lambda  \sum_{s=1}^t \varphi_s^2 
+ {1\over 8a}\left({1-\beta^2\over
1+\beta^2}\right)^2 \sum_{s=1}^t {1\over \varphi_s^2}
(\varphi_{s+1}^2-\varphi_s^2)^2,\cr} }
where both terms are of order $a$, as $\varphi_{s+1}^2 -\varphi_s^2$ is
of order $a$.
The contribution from the pre-factor $U_{\beta}(\{\alpha_s\})$ 
\sublead\ and
the measure reads 
\eqn\subsub{\prod_{s=1}^t d\alpha_s U_\beta(\{\alpha_s\}) \simeq 
\prod_{s=1}^t {\sqrt{2\over \pi a}d \varphi_s }
(1+{ a\over 8 \varphi_s^2}) 
\simeq \left(\prod_{s=1}^t {2\over\sqrt{2\pi a}}d \varphi_s\right)  
e^{{a\over 8} \sum_{s=1}^t {1\over \varphi_s^2}}.}
We notice that this latter relation has the effect of adding
an extra term to the action \actnew. 
Next, we introduce a discrete function by
$\varphi(u=s/t)\equiv \varphi_s$ and we assume that for large
$t$ (small $a$)
this function becomes a smooth function of the continuous variable 
$u\in[0,1]$. 
This allows us 
to Taylor-expand $\varphi_{s+1}$ around $\varphi_s$ as
$\varphi(u+a/T)=\varphi(u)+{a\over T}\varphi'(u)+
{\cal O}(a^2)$.
Furthermore, we can replace sums by integrals, $\sum_s\to 
{T\over a}\int_0^1
du$. Finally, it proves
convenient to define the continuum partition function $Z_T$ by
\eqn\contZ{Z_T\equiv \lim_{a\to 0} \left( {1 \over 2}
{1-\beta^2\over 1+\beta^2}\right)^{t} Z(t).}
Inserting everything and rescaling $\varphi\to \left({1+\beta^2 \over
1-\beta^2}\right)\varphi$, $u \to Tu$ we get
\eqn\funczl{ Z_T = \int_{\varphi(0)=\varphi(T)} 
{\cal D}\varphi e^{-{1\over 2}\int_0^T du 
(\varphi'(u)^2+\omega^2\varphi(u)^2 -{A\over 4\varphi(u)^2}) },}
where we have identified the functional measure as
\eqn\funcmeas{{\cal D}\varphi=\lim_{a \to 0, t={T \over a} \to \infty}
\prod_{s=1}^t {d\varphi_s \over \sqrt{2\pi a}},}
and where the integration extends over real positive  
fields $\varphi(u)$ defined for $u\in [0,T]$ and obeying
$\varphi(0)=\varphi(T)$. The parameters $\omega$ and $A$ are given by
\eqn\freq{ \omega= \sqrt{\Lambda}\,\left( {1+\beta^2 \over
1-\beta^2}\right), \qquad 
A= \left({1-\beta^2\over 1+\beta^2}\right)^2. }
Using the Feynman-Kac formula we see that computing $Z_T$ 
amounts to solving a one-dimensional (imaginary time) quantum 
mechanical system on 
the real half-line, namely
\eqn\scal{ Z_T ={\rm Tr}( e^{-T {\hat H}} ), }
with a Hamiltonian 
\eqn\hamilto{ {\hat H}= -{1\over 2} {d^2 \over d\varphi^2} 
+{1\over 2}\omega^2 \varphi^2
-{A\over 8\varphi^2},\ \ \ \ \ \ \ \varphi>0. }
This is nothing but the celebrated Calogero Hamiltonian for one
particle.
We thus arrive at the  result that $1+1$-dimensional Lorentzian
triangulations (in the low arch density regime $\beta<1$) are
equivalent in the continuum limit to a one-dimensional Calogero model.
This holds in particular for the pure case.
A Hamiltonian describing the continuum limit of pure Lorentzian 
triangulations, i.e. $\beta\to 0$
was derived in ref.~\AL\ using a strategy different from the one 
employed here. 
The Hamiltonian of ref.~\AL\ is expressed in terms of a variable $L$ which has
the interpretation of the continuum counterpart of the length $i$ of a 
line of constant time. As we 
shall see later the field $\varphi^2$ has the interpretation of the continuum
counterpart of the
total number of triangles in a given time slice $i+j$, 
also equal to the total length of the
two adjacent constant time lines, translating into $2L$ in the continuum
limit. 
Due to the use of different boundary
conditions in ref.~\AL\ and here one cannot immediately compare the two
Hamiltonians. In appendix A we show how to recover the Calogero Hamiltonian
for $\beta \to 0$ using the approach of ref.~\AL. As expected a change
of variables $2L\to \varphi^2$ is involved. 
We also explain why the approach of ref.~\AL\ does not extend to $\beta\neq 0$.

Note that the parameter $A$ of our
Hamiltonian~\hamilto\ satisfies $0\leq A\leq 1$
and that $A=1$ iff $\beta=0$. It is well-known~\LL\ that $A=1$ is a
limiting case for the quantum mechanical system~\hamilto. For $A>1$
the operator ${\hat H}$ is no longer self-adjoint. Our generalized 
Lorentzian triangulation model thus always leads to a physically acceptable
Hamiltonian. 
For $\beta <1$ the Hamiltonian ${\hat H}$ is readily diagonalized as 
follows~\LL. 
Introducing the confluent hypergeometric function
\eqn\conflu{ F(a,b,x)= \sum_{k=0}^\infty 
{a(a+1)...(a+k-1) \over k! b(b+1)...(b+k-1)} x^k,}
the eigenvectors of ${\hat H}$ take the form
\eqn\eigenH{ \psi_n(\varphi)= {\cal A}_n e^{-{1\over 2}\omega 
\varphi^2} \varphi^{\mu-{1\over 2}}
F(-n,\mu,\omega\varphi^2), }
where ${\cal A}_n$ is a normalization factor 
\eqn\valA{ {\cal A}_n=\sqrt{ 2 \omega^\mu \Gamma(n+\mu) 
\over \Gamma(n+1)\Gamma(\mu)^2}, }
ensuring the orthonormality of the eigenbasis $\psi_n(\varphi)$,
and the parameter $\mu$ may take
two values, related to the potential strength $A$ through
$\mu= 1\pm {1\over 2}\sqrt{1-A}$. 
However, the smaller of these two values must be discarded 
as unphysical~\LL, leading to
\eqn\mueq{ \mu= 1+{\beta \over 1+\beta^2}. }
Note that the confluent hypergeometric series $F$ 
is truncated by the value $a=-n$ to 
a polynomial of degree $n$, generalizing the Laguerre polynomial $L_n$,
namely 
\eqn\truc{ F(-n,\mu,z)= \sum_{k=0}^n (-z)^k {n \choose k} 
{1\over \mu(\mu+1)...(\mu+k-1)}.}
The corresponding eigenvalues read
\eqn\eigenqm{ E_n= \omega (2n+\mu) =\omega(2n+1+{\beta \over
1+\beta^2}),
\ \ \ \ \ \ \ \ n=0,1,2,...}
This leads to the scaled partition function
\eqn\scafin{ Z_T =\sum_{n\geq 0} e^{-T E_n}
={e^{-\sqrt{\Lambda} T\left({1+\beta +\beta^2 \over 1-\beta^2}\right)}\over
1-e^{-2\sqrt{\Lambda} T\left({1+\beta^2\over 1-\beta^2}\right)} }.}
For $\beta \to 0$ the result \scafin\ reduces to 
the pure Lorentzian triangulation result \Zcont. 
On the level of eigenvalues and eigenfunctions we have
$\lim_{a\to 0}(g \lambda_n)^{t}=\lim_{\mu\to 1} e^{-E_n T}$ and
$\lim_{a\to 0} \psi_n^{(0)}(\alpha) \sqrt{d \alpha} = \lim_{\mu\to 1}
\psi_n(\varphi) \sqrt{d \varphi}$,
respectively obtained from
eqns. \psinordef\ and \eigenH\ with $\omega\to \sqrt{\Lambda}$,
and after setting
$q=e^{-a\sqrt{\Lambda}}$, $t={T \over a}$ and $\alpha={\varphi^2 \over a}$
as before.

Furthermore, we see that as
$\beta \to 1$ the exponents in \scafin\ become singular, signaling another type
of scaling coming into play (c.f. Sect. 4.2 and 4.3 above). 
It is worth stressing that the parameter $\beta$ cannot be absorbed by a
redefinition of the renormalized triangle fugacity $\Lambda$ or the 
continuum time variable, and is itself a genuine continuum parameter. 
This will prove even more visible
when we compute correlation functions. 
Such a situation where
a coupling constant introduced at the discrete level survives in an
unrenormalized form in the continuum limit is quite unusual.

\subsec{The loop-loop correlator}

The Laplace transform of the loop-loop correlator $Z(x,y,t)$ can be
written as (cf.\ eqn.~\Zij)
\eqn\geneloop{\eqalign{
Z(x,y,t)=&\sum_{i,j\geq 0} Z(i,j,t)\,x^i y^j \cr
=& \prod_{s=1}^t\bigg(\oint_{\cal C}{d w_s \over 2i\pi w_s}
{d z_s \over 2i\pi z_s}\bigg) \times \cr
&\Theta^{(1)}\left(x,{1\over w_1}\right)
\left( \prod_{r=1}^{t-1} 
\Theta^{(2)}(w_r,z_r)\Theta^{(1)}\left({1\over z_r},{1\over 
w_{r+1}}\right)\right)
\Theta^{(2)}(w_t,z_t)\Theta^{(1)}\left({1\over z_t},y\right). \cr}}
Repeating the computation of Sect.\ 3, namely performing
the change of variables $w_r,z_r\to \mu_r,\rho_r$ (cf.\
eqns.~\chvar\ and \invch),  representing
the $\Theta^{(1)}$-factors as Schwinger-type integrals, and
performing the explicit contour integrals over the $\rho_r$
and $\mu_r$, it appears that $x$ and $y$ are simply spectators throughout 
the computation, and one finds
\eqn\endcomp{ Z(x,y,t)= 
\prod_{s=0}^t 
\left(\int_0^\infty 
d\alpha_se^{-\alpha_s} \right) e^{g (\alpha_0 x+\alpha_t y)} 
\prod_{r=0}^{t-1} \phi_\beta(g \theta \alpha_r,g \theta \alpha_{r+1}),}
where the transition function $\phi_\beta(x,y)$ has been defined in
\transit.
Apart from the explicit exponential dependence on $x$ and $y$,
the loop-loop correlator differs from the partition function~\lutfin\
by involving one more Schwinger parameter.
Note also that in this case we do not impose periodic boundary
conditions on the $\alpha$'s.

We will now study the behavior of the loop-loop correlator in the
scaling limit defined by~\betascal\ and~\Tcontbeta. 
Here we must also assume
that $x$ and $y$ are close to their critical values $x_c=y_c={1\over 2g}$ 
(corresponding to the poles of the $\Theta^{(1)}$-factors in \geneloop).
It turns out that the correct scaling ansatz reads
\eqn\scanz{ x= {1\over 2g} ( 1- a X), \qquad 
y={1\over 2g}(1- a Y).}
To calculate $Z(x,y,t)$ in the scaling limit we use the same strategy
as in Sect.~5.1. We perform the change of variables $\alpha_s={\varphi_s^2
\over a}$, $s=0,1,\ldots,t$, introduce a discrete function by
$\varphi(u=s/t)\equiv\varphi_s$ and assume that as $t\to
\infty$ this function becomes a smooth function of a continuous
variable $u\in [0,1]$. However, in the present case we must deal with
fixed as opposed to periodic boundary conditions for $\varphi$ and
special care has to be taken in the treatment of the boundary terms.
Writing the integrand in~\endcomp\ for large $\alpha_s$ as
$e^{-S_{\beta}} U_{\beta}$ in the same way
as in Sect.~3, we get, when inserting our change of variables,
the following contribution
from the measure and the pre-factor
\eqn\subfix{ \big(\prod_{s=0}^t d\alpha_s\big)
{U}_\beta(\{\alpha_s\}) = {2^{t+1}
\over a} {1\over \sqrt{2 \pi a}}
d\varphi_0 \sqrt{\varphi_0}
d\varphi_t \sqrt{\varphi_t}
\big(\prod_{s=1}^{t-1} {d \varphi_s\over \sqrt{2 \pi a}}\big)
e^{{a\over 8}\big({1\over 2}({1\over \varphi_0^2}+{1\over \varphi_t^2})
+\sum_{r=1}^{t-1}{1\over \varphi_r^2}\big) +{\cal O}(a^2)}.}
Inserting the various scaling relations in $S_{\beta}(\{\alpha_s\})$
on sees that the scaling ansatz~\scanz\ is exactly what is needed to
make divergent boundary terms cancel. Furthermore it becomes 
natural to define a continuum loop-loop correlator $Z_T(X,Y)$ by the
following recipe
\eqn\contprop{Z_T(X,Y)=\lim_{a\to 0} a\cdot {1 \over 2^{t+1}} 
\left({1-\beta^2 \over 1+\beta^2}\right)^t Z(x,y,t).}
Here we choose to scale away the same entropic factor as in the case
of partition function, namely a factor ${2}$ for each occurrence of
the transfer matrix $\Theta^{(1)}$ and a factor $\left({1+\beta^2\over
1-\beta^2}\right)$ for each occurrence of the transfer matrix
$\Theta^{(2)}$. 
Collecting all the terms and rescaling $\varphi\to \left({1-\beta^2 \over
1+\beta^2}\right)\varphi$ and $u\to Tu$ as before, we find that the continuum
loop-loop correlator can be expressed as the quantum mechanical
propagator
\eqn\qmprop{ Z_T(X,Y)= 
\left( {1+\beta^2\over 1-\beta^2}\right)^{2} 
\int_0^\infty \sqrt{\varphi_0}d\varphi_0 
\sqrt{\varphi_T}d\varphi_T 
e^{-{1\over 2}(\varphi_0^2 {\tilde X}+\varphi_T^2 {\tilde Y})} 
\langle \varphi_T \vert e^{-T {\hat H}}\vert \varphi_0\rangle,}
where the Hamiltonian ${\hat H}$ is defined as in \hamilto, 
with $\omega$ and $A$ as in \freq, and where we have set
\eqn\setXY{ {\tilde X} = \left({1+\beta^2\over 1-\beta^2}\right)^2 X,
\qquad {\tilde Y} = \left({1+\beta^2\over 1-\beta^2}\right)^2 Y.}
To evaluate $Z_T(X,Y)$ 
we simply have to insert two decompositions of the identity 
as a sum over projectors on the eigenspaces of ${\hat H}$, i.e.\
$I=\sum_{n\geq 0} \vert n\rangle
\langle n\vert$  and use 
$\langle n \vert \varphi\rangle= \psi_n(\varphi)$, 
where $\psi_n(\varphi)$ is the $n$-th
normalized eigenfunction of ${\hat H}$, defined in \eigenH. 
This yields
\eqn\eigenpaquet{\eqalign{ Z_T(X,Y)&=\left({1+\beta^2 \over
1-\beta^2}\right)^2 \sum_{n=0}^\infty 
e^{-T E_n} G_n({\tilde X}) G_n({\tilde Y}),\cr
G_n(Z)&=\int_0^\infty \sqrt{\varphi}d\varphi 
e^{-{1\over 2}\varphi^2 Z} \psi_n(\varphi), \cr}}
with $E_n$ as in \eigenqm.
The function $G_n(Z)$ is readily determined using the expression 
\eigenH\ for $\psi_n(\varphi)$ 
\eqn\evalF{G_n(Z)={\cal A}_n\int_0^\infty d\varphi \varphi^{\mu}
e^{-{1\over 2}\varphi^2 (\omega+Z)} F(-n,\mu,\omega \varphi^2), }
where $\mu=1+{\beta\over 1+\beta^2}$ as before.
Inserting the polynomial expression for the truncated confluent
hypergeometric function \truc,
eqn. \evalF\ can be integrated term by term with the result
\eqn\resF{\eqalign{ G_n(Z)&={{\cal A}_n\over 2} \left({2\over \omega+Z}
\right)^{\mu+1\over 2} \sum_{k=0}^n
{n \choose k} {\Gamma({\mu+1\over 2}+k)\over \mu(\mu+1)...(\mu+k-1)}
\left(-{2 \omega \over \omega+Z}\right)^k \cr
&={{\cal A}_n\over 2} \Gamma({\mu+1\over 2}) 
\left({2\over \omega+Z}\right)^{\mu+1\over 2}
F(-n,{\mu+1\over 2},\mu,{2 \omega\over \omega+Z}). \cr}}
Here we have recognized the ordinary hypergeometric function
\eqn\hypergeometric{
F(a,b,c,z) \equiv _2\!\! F_1(a,b,c,z)
=\sum_{k=0}^\infty z^k{a(a+1)...(a+k-1) b(b+1)...(b+k-1)\over
k!\ c(c+1)...(c+k-1)}.}
Thus, the continuum loop-loop correlator reads
\eqn\scallop{\eqalign{
 Z_T&(X,Y)=\left({1+\beta^2\over 1-\beta^2}\right)^2
{1 \over 2 \omega} 
\left({\Gamma({\mu+1\over 2})\over \Gamma(\mu)}\right)^2 
\left({4\omega^2\over (\omega+{\tilde X})(\omega+{\tilde Y})} 
\right)^{\mu+1\over 2} \cr
&\times \sum_{n=0}^\infty
e^{-\omega T (2n+\mu)} {\Gamma(n+\mu)\over \Gamma(n+1)} 
F(-n,{\mu+1\over 2},\mu,{2\omega\over \omega+{\tilde X}}) 
F(-n,{\mu+1\over 2},\mu,{2\omega\over \omega+{\tilde Y}}).\cr} }
This expression may further be simplified by use of the 
following quadratic relation due to Meixner \BATI\ 
\eqn\meix{\eqalign{ \sum_{n=0}^\infty & {\Gamma(\mu+n)\over \Gamma(\mu) n!}
(-s)^n F(-n,a,\mu,z)F(-n,b,\mu,w) \cr
&= {(1+s)^{a+b-\mu}\over
(1+s(1-z))^a (1+s(1-w))^b} F(a,b,\mu,-{s z w\over (1+s(1-z))(1+s(1-w))}).\cr}}
Indeed, applying \meix\ with  
$s=-e^{-2T \omega}$, $a=b={\mu+1\over 2}$, 
$z=2\omega/(\omega+{\tilde X})$, and $w=2\omega/(\omega+{\tilde Y})$, 
the continuum loop-loop correlator finally reads
\eqn\lopcor{Z_T(X,Y)=\left({1+\beta^2 \over 1-\beta^2}
\right)^2
{\sinh(\omega T)\over \omega} {\Gamma({\mu+1\over 2})^2\over \Gamma(\mu)} 
v^{\mu+1\over 2} F({\mu+1\over 2},{\mu+1\over 2},\mu,v),}
where
\eqn\v{v={\omega \over
({\tilde X} \sinh(\omega T) +\omega \cosh(\omega T))} \times {\omega \over
({\tilde Y} \sinh(\omega T) +\omega \cosh(\omega T)) } .}

Let us consider the pure Lorentzian triangulation limit $\beta\to 0$, 
i.e.\ $\mu\to 1$ and $\omega\to\sqrt{\Lambda}$. This limit involves
the hypergeometric function $F(1,1,1,v)=1/(1-v)$, and the corresponding 
loop-loop propagator reads
\eqn\loopure{\eqalign{ Z^{(0)}_T(X,Y)&= 
{\sinh (\omega T)\over \omega}\,\left. {v \over
1-v}\right|_{\omega=\sqrt{\Lambda}}
\cr
&=
{\sqrt{\Lambda} \over  
(XY+\Lambda)\sinh(\sqrt{\Lambda} T)+
\sqrt{\Lambda}(X+Y)\cosh(\sqrt{\Lambda} T)}.\cr}} 
This coincides precisely with eqn. (2.29) of ref.~\FGK.

Let us now translate the result \lopcor\ into the language of conjugate
rescaled loop lengths by setting
\eqn\rescaloop{
L_0  = a i, \qquad L_T = a j, \qquad T=a t, }
in terms of which the rescaled loop-loop correlator reads
\eqn\loopresca{ Z_T(L_0,L_T)=\lim_{a\to 0} {1 \over a}{1\over 2^{t+1}} 
\left({1-\beta^2\over 1+\beta^2} \right)^t \, Z(i,j,t). }
This is obtained by taking 
the inverse Laplace transform of \lopcor\ over $X$ and $Y$. 
In practice, it is more
convenient to read it directly off the original expression 
\qmprop\ upon performing the change of variables
\eqn\changb{ L_0 = {1\over 2} \varphi_0^2 \bigg({1+\beta^2\over
1-\beta^2}\bigg)^2, \qquad 
L_T = {1\over 2} \varphi_T^2 \bigg({1+\beta^2\over
1-\beta^2}\bigg)^2,}
relating $\varphi_0^2$ resp. $\varphi_T^2$ to the numbers
of triangles pointing up resp.\ down in the time slice $u=0$ resp. $u=T$.
This yields
\eqn\resloplap{
Z_T(L_0,L_T)=\left({1-\beta^2\over 1+\beta^2}\right)^2
\left. {\langle \varphi_T\vert
e^{-T{\hat H}} \vert \varphi_0 \rangle\over 
\sqrt{\varphi_0\varphi_T}}\right\vert_{\varphi_0={1-\beta^2\over  
1+\beta^2}\sqrt{2L_0},\varphi_T={1-\beta^2\over 
1+\beta^2}\sqrt{2L_T}} }
To evaluate the above heat kernel, we can 
use the following quadratic
relation satisfied by the generalized Laguerre polynomials introduced
in Sect. 5.1, and generalizing the pure case \meixL, namely \BAT\
\eqn\laquad{\sum_{n\geq 0} {\Gamma(n+\mu)\over n! \Gamma(\mu)^2}
F(-n,\mu,x)F(-n,\mu,y) z^n ={1\over 1-z} e^{-{z(x+y)\over 1-z}}
(xyz)^{1-\mu\over 2} I_{\mu-1}(2 {\sqrt{xyz}\over 1-z}) , }
where $I_{\mu-1}(2x)= \sum_{p\geq 0} x^{\mu-1+2p}/(p!\Gamma(p+\mu))$
is the modified Bessel function.
This equation is immediately rephrased in terms of the eigenfunctions
$\psi_n$ of $\hat H$ \eigenH\ as
\eqn\noybet{
\sum_{n\geq 0} z^{n+{\mu\over 2}} \psi_n(\varphi_0)
\psi_n(\varphi_T) = 2 \omega \sqrt{\varphi_0\varphi_T} {\sqrt{z}\over 1-z}
e^{-{1+z\over 1-z}{\omega\over 2}(\varphi_0^2+\varphi_T^2)}
I_{\mu-1}\left(2 {\sqrt{z}\omega\varphi_0\varphi_T\over 1-z}\right).}
Taking $z=e^{-2 \omega T}$ we deduce the expression for the heat kernel
of the Calogero Hamiltonian $\hat H$, leading finally to
\eqn\lutfinlop{\eqalign{
Z_T(L_0,L_T)=\left({1-\beta^2\over 1+\beta^2}\right)^2
&{\omega \over {\rm sinh}(\omega T)} e^{-(L_0+L_T)\big({1-\beta^2\over
1+\beta^2}\big)^2 \omega {\rm cotanh}(\omega T)} \times \cr 
&\times I_{\mu-1}\left( 2\bigg({1-\beta^2\over 1+\beta^2}\bigg)^2
{\omega \over {\rm sinh}(\omega T)} 
\sqrt{L_0L_T}\right),\cr} }
with $\omega$ as in \freq\ and $\mu-1=\beta/(1+\beta^2)$. This result
coincides exactly with eqn. (2.32) of~\FGK.

\subsec{Correlation functions on a time cylinder}

In this section, we will compute for $\beta<1$ the general 
correlations \corsig\ of the
observables $\Sigma_{s_m}(z_m)$ \factomo, in the continuum limit where
$a \to 0$, with 
\eqn\contbeto{ s_m= {u_m \over a} \qquad ,
\qquad z_m = a \omega {\cal Z}_m \qquad  {\rm and} 
\qquad \alpha_{s_m}= {\varphi^2(u_m)\over a}. }
With these substitutions, eqn. \corsig\ becomes
\eqn\sigcor{ \langle \prod_{m=1}^k e^{\omega {\cal N}(u_m) {\cal Z}_m}\rangle
= \langle \prod_{m=1}^k e^{\omega \varphi^2(u_m) {\cal Z}_m}\rangle , }
where the rescaled number of triangles in the time-slice $u\in [0,T]$ reads
${\cal N}(u)=a\big({1-\beta^2\over 1+\beta^2}\big)^2 N(u/a)$. 
Apart from its instrumentality, eqn. \sigcor\
yields the interpretation of the field $\varphi(u)$: the square of $\varphi(u)$
is identified with the rescaled number ${\cal N}(u)$.
This is very reminiscent of the field theoretical representation of  
polymers, where $\varphi^2$ is identified with the polymer density. 
For the pure case, this should not come as a surprise as we 
have shown in \FGK\ that Lorentzian triangulations can be mapped onto
random walks.

Let us first compute the one-point average of $e^{\omega {\cal N} {\cal Z}}$: 
\eqn\oneptav{ f_\mu({\cal Z})\equiv\langle e^{\omega {\cal N} {\cal Z}}\rangle=
{1\over Z_{T,\mu}} \int_0^\infty d\varphi
\sum_{n\geq 0} \psi_n(\varphi)^2 e^{-\omega T(2 n+\mu)} 
e^{{\cal Z}\omega\varphi^2},}
with the cylinder partition function
$Z_{T,\mu}=e^{-\omega T \mu}/(1-e^{-2 \omega T})$, indexed by 
$\mu=1+\beta/(1+\beta^2)$ for convenience.
We use again the above quadratic equation \noybet\ with $z=e^{-2 \omega T}$
and $\varphi_0=\varphi_T=\varphi$, and we perform the change of variables
$u\equiv\omega\varphi^2$ to get
\eqn\toget{ f_{\mu}({\cal Z})=e^{\omega T(\mu-1)} \int_0^\infty du  e^{-({\rm
cotanh}(\omega T)-{\cal Z})u}
I_{\mu-1}\left({u \over {\rm sinh}(\omega T)}\right) .}
This last integral is a particular case of the generic Laplace transform
of the modified Bessel \BAT:
\eqn\useful{
r \int_0^\infty du e^{-r u} I_{\mu-1}(s u)={(r/s)^{\mu-1} \over 
\sqrt{1-(s/r)^2}}
(1-\sqrt{1-(s/r)^2})^{\mu-1}, }
and we finally get
\eqn\oneptfin{f_{\mu}({\cal Z})=
{\left( {{\rm cotanh}(\omega T)-{\cal Z} -\sqrt{1-2 {\rm cotanh}(\omega
T) 
{\cal Z}+{\cal Z}^2}
\over {\rm cotanh}(\omega T)-1}\right)^{\mu-1}
\over \sqrt{1-2 {\rm
cotanh}(\omega T){\cal Z} +{\cal Z}^2}} ,}
generalizing the pure case result \scacozer\ corresponding to $\mu=1$ and
$\omega=\sqrt{\Lambda}$. 
This yields alternatively the moments $\langle {\cal N}^k\rangle$ of the
rescaled number of triangles per time-slice 
\eqn\jacobi{ \langle {\cal N}^k\rangle = {k! \over \omega^k} 
P_k^{(\mu-1,1-\mu)}({\rm cotanh}(\omega T))=
{k! \over \omega^k} {\sum_{m=0}^k {k+\mu-1 \choose m} {k+1-\mu\choose
k-m} e^{-2 m \omega T} \over (1-e^{-2\omega T})^k }, }
where $P_k^{(\alpha,\beta)}(x)$ denotes the $k$-th Jacobi polynomial \BAT.

The formula \oneptfin\ generalizes nicely to the case of the $k$-point
function \sigcor, and we leave the details of its derivation to appendix B
below. Defining the time intervals
\eqn\intervals{ t_m=u_{m+1}-u_m, }
for $m=1,2,...,k-1$ and $t_k=T-\sum_{1\leq m\leq k-1} t_m$, 
the result reads
\eqn\kptfin{  
\langle \prod_{m=1}^k e^{\omega {\cal N}(u_m) {\cal Z}_m}\rangle
=g_\mu\big({2 \over P({\cal Z}_1,...,{\cal Z}_k \vert t_1,...,t_k)}\big),}
where
\eqn\defgmu{g_\mu(s)={\rm sinh}(\omega T) e^{\omega T(\mu-1)}
s^{2-\mu} {(1-\sqrt{1-s^2})^{\mu-1}\over \sqrt{1-s^2}}=f_\mu\big({\rm
cotanh}(\omega T)-{1\over s \, {\rm sinh}(\omega T)}\big) ,}
with $f_\mu$ as in \oneptfin,
and where
$P$ is the following polynomial of ${\cal Z}_1$, ..., ${\cal Z}_k$ 
\eqn\Ppol{\eqalign{ P&({\cal Z}_1,...,{\cal Z}_k\vert t_1,...,t_k)
=2 {\rm cosh}( \omega T) \cr
&+\sum_{r=1}^k (-1)^r
\sum_{1\leq m_1 <m_2 <...<m_r\leq k} {\cal Z}_{m_1} {\cal Z}_{m_2} ...
{\cal Z}_{m_k}
\prod_{j=1}^r 2\, {\rm sinh}\big(\omega(\sum_{m_j\leq s\leq m_{j+1}-1}
t_{s})\big) , \cr}}
with suitable boundary conditions on the indices, namely: 
$m_{r+1}\equiv m_1+k$ and $t_s\equiv t_{s-k}$ for $s>k$.
Note the remarkable fact that all the $k$-point correlators
are expressed in terms of the same universal scaling function $g_\mu$ 
containing all the $\mu$-dependence.

Moreover, the results of this section clearly only depend on the
parameters $\mu,\omega,T$, so forgetting about the dependence of $\mu$
and $\omega$
on $\beta$, we may in particular interpret the cases $\mu=1/2$ and 
$\mu=3/2$ in terms of a one-dimensional harmonic oscillator.
Indeed, when $\mu=1/2$ (resp. $\mu=3/2$), 
our quantum system reduces to the even (resp. odd) sector of
an ordinary harmonic oscillator with frequency $\omega$ on a time circle
of length $T$. 
More precisely, the eigenvalues
$E_n=\omega(2n+\mu)$ \eigenqm\ and eigenvectors
$\psi_n$ \eigenH\ reduce respectively to
\eqn\respred{\eqalign{
\mu={1\over 2}\ : \ \ \ &E_n=\omega(2n+{1\over 2}),\qquad 
\psi_n(\varphi)= {(-1)^n\over 2^{n-{1\over 2}} \sqrt{(2n)!}}
\big({\omega\over \pi}\big)^{1\over 4} H_{2n}(\sqrt{\omega}\varphi)
e^{-{1\over 2}\omega \varphi^2} , \cr 
\mu={3\over 2}\ : \ \ \ &E_n=\omega(2n+1+{1\over 2}),\qquad 
\psi_n(\varphi)={(-1)^n\over 2^{n} \sqrt{(2n+1)!}}
\big({\omega\over \pi}\big)^{1\over 4}  H_{2n+1}(\sqrt{\omega}\varphi)
e^{-{1\over 2}\omega \varphi^2}, \cr}} 
in terms of the Hermite polynomials $H_n(x)$.
Note that $\varphi$ is still restricted to be positive, but 
we may relax this condition by remarking that the Hermite polynomials of
even degree are even, while the odd degree ones are odd. 
So we may extend the range of $\varphi$ to 
the whole real line, upon simply redefining the eigenvectors above $\psi_n\to
\psi_n/\sqrt{2}$, and considering only even observables, say. 
The corresponding partition functions read
respectively
\eqn\parhar{ Z_{T,{1\over 2}}= {e^{-{\omega T\over 2}} \over
1-e^{-2\omega T}} \qquad {\rm and} \qquad 
Z_{T,{3\over 2}}= {e^{-3{\omega T\over 2}}
\over 1-e^{-2\omega T}} , }
and their sum is nothing but the partition function of the 
harmonic oscillator
\eqn\harmo{Z_{T,\rm osc}=Z_{T,{1\over 2}}+Z_{T,{3\over 2}}
={e^{-{\omega T\over 2}} \over 1-e^{-\omega T}} .}
More generally, the result \kptfin\ for the correlation functions of the
operator $e^{\omega{\cal N}{\cal Z}}$
can be used to derive the corresponding quantity for the harmonic oscillator.
Indeed, the odd and even sectors are orthogonal, and the full correlator reads
\eqn\harmoco{\eqalign{
\langle \prod_{m=1}^k &e^{\omega\varphi^2(u_m){\cal Z}_m} \rangle_{\rm osc}=
{1\over Z_{T,\rm osc}}( Z_{T,{1\over 2}} g_{1\over 2}(s) + Z_{T,{3\over 2}} 
g_{3\over 2}(s) ) \cr
&={\rm sinh}\left({\omega T\over 2}\right)\ 
\sqrt{s \over 1-s^2}\ ( s (1-\sqrt{1-s^2})^{-{1\over 2}} +
(1-\sqrt{1-s^2})^{1\over 2}) \cr
&= {\rm sinh}\left({\omega T\over 2}\right)\ \sqrt{2s \over 1-s},\cr} }
where $s=2/P({\cal Z}_1,...,{\cal Z}_k\vert t_1,...,t_k)$ and $g_\mu$ as in
\defgmu. From \harmoco\ we learn that the corresponding correlation
function in the harmonic oscillator case also only depends on the arguments
of the observables through the quantity $s$. In Appendix C, we compute
directly the correlator \harmoco\ within the framework of the harmonic 
oscillator
and obtain an alternative expression for the combined argument $s$, which
leads to the following determinantal expression for the polynomial $P$
\eqn\deterexp{ P({\cal Z}_1,...,{\cal Z}_k\vert t_1,...,t_k)=2\big(
1+ 2 {\rm sinh}^2({\omega T\over 2}) \ {{\rm det}(M-Z)
\over {\rm det}(M)} \big), }
where $M$ and $Z$ are the following $k\times k$ matrices
\eqn\matrimz{
M=\pmatrix{ \gamma_1 & -x_1  & 0 & \cdots & -x_k\cr
-x_1 & \gamma_2 & -x_2 & \cdots & 0 \cr
0 & -x_2 & \gamma_3 & \cdots & \vdots \cr
\vdots & \ddots & \vdots & \ddots & \vdots \cr
-x_k & 0 & \cdots & -x_{k-1} & \gamma_k \cr}
 \qquad  
Z=\pmatrix{ {\cal Z}_1 & 0 & 0 & \cdots & 0 \cr
0 & {\cal Z}_2 & 0 & \cdots & 0 \cr
0 & 0 & {\cal Z}_3 & \cdots & \vdots \cr
\vdots & \ddots & \vdots & \ddots & \vdots \cr
0 & \cdots & \cdots & 0 & {\cal Z}_k \cr}, }
with $x_i=1/(2 {\rm sinh}( \omega t_i))$, and $\gamma_i=2 x_ix_{i-1}
{\rm sinh}(\omega(t_i+t_{i-1}))$, $i=1,2,...,k$, 
with the convention $x_0\equiv x_k$, $t_0\equiv t_k$.

\newsec{Discussion}

\subsec{Lorentzian vs.\ Euclidean triangulations}

With the present work we have added yet another case to the list of problems
for which the combination of methods from quantum field theory and
statistical mechanics has proven very powerful (see for instance~\ADJ,\ID).

We have seen that in a certain region of the coupling constant space of our
model, more precisely for $\beta<1$, it is possible to define a continuum
limit. The coupling constant $\beta$ survives this limit in unrenormalized
form and we get a one-parameter family of continuum models which are neither
identical to the continuum description of pure Lorentzian triangulations nor 
to that of Euclidean
ones. The models do, however, have some common features with 
the pure Lorentzian triangulation model: they have the same scaling of the 
time
variable and thus the same fractal dimension $d_H=2$ of the continuum 
space-time manifolds.
One could have hoped, bearing in mind our
original aim at gaining a better understanding of the renormalization idea
of~\ACKL, that one would get out Euclidean quantum gravity for
$\beta=1$. However, we are led to the conclusion that no continuum limit
exists for $\beta\geq 1$. The class of outgrowths that we considered
here is too restricted 
to allow the model to flow from the Lorentzian case to
the Euclidean one. What happens instead as $\beta$ increases from zero to
one is that the correlation length $\xi$ defined by (cf.\ eqn. \scafin)
\eqn\corlength{Z_T\sim \exp\left(-{1\over \xi} T\right), \ \ \ \ \  
\rm {as}\ \ \ \ \ T \to \infty ,}
i.e.
\eqn\xxi{\xi=
\left({1-\beta^2 \over 1+\beta+\beta^2}\right){1 \over \sqrt{\Lambda}},} 
decreases from ${1\over \sqrt{\Lambda}}$ to zero. This means that our time
slices effectively decouple from each other, not allowing us to
interpret our model as a model of surfaces. This decoupling effect is of
course due to the fact that the outgrowths which dominate the 
geometries more and more
as $\beta$ increases do not imply any interaction between different
time slices. 

\subsec{Effective integrable structure and Renormalization Group formulation}

In ref. \FGK, a two-parameter family of theories has already been introduced
to describe Lorentzian triangulations with intrinsic curvature energy 
(the parameters
being the weight per triangle and a curvature weight). 
These models were found to be integrable, in the sense that their
transfer matrices taken at different values of the curvature weight 
and of the weight per triangle but along the same spectral curve 
commute with one another. 
However, all these models turned out to share the {\it same} continuum
limit up to a redefinition of the space and time scales which absorbed
the (irrelevant) curvature parameter.
One of those models is precisely the pure case $\beta=0$. 

Here we have a very different situation: the parameter $\beta$ survives
and leads to distinct continuous theories. On the other hand, the 
model we started from does not seem to be part of an integrable family
(except for $\beta=0$). Still, the very simple form of the energies
\eigenqm\ suggests that for each value of $\beta<1$ there exists a
two-parameter family of transfer matrices which is integrable and
leads to the same continuum limit as our original model. 
The corresponding kernels are readily constructed by use of the quadratic
relation \noybet. 
Upon taking
\eqn\takeint{ z=\lambda,
\qquad \varphi_0=\sqrt{\alpha}, \ \varphi_T=\sqrt{\alpha'}, }
we may introduce the kernel
\eqn\droite{\eqalign{
G_{\Omega,\lambda}(\alpha,\alpha')
&\equiv\sum_{n\geq 0} {\psi_n^{(\Omega)}(\sqrt{\alpha})\over \sqrt{2 
\sqrt{\alpha}}}
{\psi_n^{(\Omega)}(\sqrt{\alpha'})\over \sqrt{2 \sqrt{\alpha'}}} 
\lambda^{n+{\mu\over 2}}\cr
&= {\Omega\sqrt{\lambda}\over 
1-\lambda} \ e^{-{1\over 2}{1+\lambda \over 1-\lambda}
\Omega (\alpha+\alpha')} \ I_{\mu-1}\left(2\Omega {\sqrt{\lambda}
\over 1-\lambda} \sqrt{\alpha\alpha'}\right) ,\cr }}
where $\mu-1=\beta/(1+\beta^2)$ as before
and the $\psi_n^{(\Omega)}$ are given by 
\eigenH\ with the substitution $\omega\to \Omega$.
By construction, it is clear that
\eqn\intrint{ \int_0^\infty d\alpha'' G_{\Omega,\lambda}(\alpha,\alpha'')
G_{\Omega,\lambda'}(\alpha'',\alpha') =
G_{\Omega,\lambda\lambda'}(\alpha,\alpha'),}
hence all the transfer kernels commute at fixed $\Omega$.
Choosing
\eqn\setOmeg{ \Omega= \Omega(\lambda)\equiv {1-\lambda\over 1+\lambda} 
\left({1-\beta^2\over 1+\beta^2}\right)^2, }
and rescaling the $\alpha$'s as 
\eqn\recalph{ {\tilde \alpha} = \left({1-\beta^2\over 1+\beta^2}\right)^2 
\alpha,
\qquad {\tilde \alpha}' = \left({1-\beta^2\over 1+\beta^2}\right)^2 \alpha',}
we get a one-parameter family of kernels
\eqn\newparam{ {\tilde G}_\lambda({\tilde \alpha},{\tilde \alpha}')\equiv
\left({1+\beta^2\over 1-\beta^2}\right)^2\, 
G_{\Omega(\lambda),\lambda}(\alpha,\alpha')=
{\sqrt{\lambda}\over 1+\lambda}
e^{-{1\over 2}({\tilde \alpha}+{\tilde \alpha}')}
I_{\mu-1}\left(2 {\sqrt{\lambda}\over 1+\lambda} 
\sqrt{{\tilde \alpha}{\tilde \alpha}'}\right).}
The kernels \newparam\ lead to the same continuum limit
$a\to 0$ as our original generalized Lorentzian triangulation model with 
parameters $\beta$ and $g\theta$
upon taking
\eqn\upontak{ \lambda=e^{-2 a \omega}=e^{-2 a \sqrt{\Lambda}\big({1+\beta^2\over
1-\beta^2}\big)},  }
with the same $a$ and $\Lambda$ as in \betascal.

We may now reinterpret the composition formula \intrint\ as a self-similarity
property for the kernel ${\tilde G}_\lambda$. Indeed, taking
$\lambda=\lambda'$, we get
\eqn\selfsim{\eqalign{\sqrt{d{\tilde \alpha} d{\tilde \alpha'}} 
\int_0^\infty d{\tilde \alpha''}
{\tilde G}_{\lambda}({\tilde\alpha},{\tilde \alpha''})  
&{\tilde G}_{\lambda}({\tilde \alpha''},{\tilde \alpha'})=
\sqrt{d\alpha d\alpha'}
G_{\Omega(\lambda),\lambda^2}(\alpha,\alpha') \cr
&={\lambda\over (1+\lambda)^2}
\sqrt{d{\tilde \alpha} d{\tilde \alpha'}}\  e^{-{1\over 2}{1+\lambda^2\over
(1+\lambda)^2}({\tilde \alpha}+{\tilde \alpha'})} 
I_{\mu-1}\left(2 {\lambda\over (1+\lambda)^2}
\sqrt{{\tilde \alpha}{\tilde \alpha'}}\right) \cr
&={\lambda \over 1+\lambda^2} \sqrt{d{\hat \alpha} d
{\hat \alpha}'} \ e^{-{1\over 2}({\hat \alpha}+{\hat \alpha'})} 
I_{\mu-1}\left(2{\lambda\over 1+\lambda^2}\sqrt{{\hat \alpha}
{\hat \alpha}'}\right) \cr
&=\sqrt{d{\hat \alpha} d
{\hat \alpha}'} {\tilde G}_{\lambda^2}({\hat \alpha}
,{\hat \alpha}'), \cr}}
where we have performed the change of variables ${\hat \alpha}=
{1+\lambda^2\over (1+\lambda)^2}\, {\tilde \alpha}$ 
and similarly on $\alpha'$, and as before ${\tilde \alpha}=
\big({1-\beta^2\over 1+\beta^2}\big)^2 \alpha$. 
The operation performed in \selfsim\ decomposes into two steps: (i)
composition of ${\tilde G}_\lambda$ with itself, by integration over
the intermediate variable ${\tilde \alpha''}$ (ii) 
rescaling of the variables ${\tilde \alpha},{\tilde \alpha'}\to {\hat
\alpha}{\hat \alpha'}$, to yield the kernel ${\tilde
G}_{\lambda^2}$.
This is nothing but a real-space functional renormalization group (RG) 
decimation procedure. Indeed,
imagine we wish to evaluate a product of say $2^n$ identical kernels 
${\tilde G}_\lambda$. 
We first apply \selfsim\ to the $2^{n-1}$ consecutive pairs of kernels, and
end up after rescaling with a product of $2^{n-1}$ kernels 
${\tilde G}_{\lambda^2}$, with the renormalized value $\lambda^2$ of the
parameter. 
Iterating the
process will lead to the final single kernel 
${\tilde G}_{\lambda^{2^n}}(\alpha_n,\alpha_n')$,
where 
\eqn\iteralph{ \alpha_n = {1+\lambda^{2^n} \over (1+\lambda^{2^{n-1}})^2}
\alpha_{n-1}, \qquad \alpha_0={\tilde \alpha}, }
and similarly for $\alpha_n'$. This is easily solved as
\eqn\varalph{ \alpha_n= {1+\lambda^{2^n} \over 1-\lambda^{2^n}} 
{1-\lambda \over 1+\lambda} \ {\tilde \alpha}, }
and similarly for $\alpha_n'$.
This yields an alternative derivation of the loop-loop correlator \loopresca\ 
over a time lapse $t=2^n$. Setting $\lambda=e^{-2 a \omega}$, $t=2^n=T/a$ 
as usual and taking the initial boundary values
\eqn\bouval{ {\alpha}_0={\tilde \alpha}=2\bigg({1-\beta^2\over  
1+\beta^2} \bigg)^2 \, {L_0\over a}, \qquad
{\alpha}_0'={\tilde \alpha}'=2\bigg({1-\beta^2\over 
1+\beta^2} \bigg)^2 \, {L_T\over a},}
we get the iterated values
\eqn\iterval{ {\alpha}_n=2\omega \, {\rm cotanh}(\omega T)
\bigg({1-\beta^2\over 
1+\beta^2} \bigg)^2 L_0, \qquad
{\alpha}_n'=2\omega \, {\rm cotanh}(\omega T)
\bigg({1-\beta^2\over 
1+\beta^2} \bigg)^2 L_T,}
so that the loop-loop correlator reads
\eqn\lolop{\eqalign{ Z_T&(L_0,L_T)= 2 \ \bigg({1-\beta^2\over
1+\beta^2}\bigg)^2 \omega \  {\rm cotanh}(\omega T)  \ \times \cr
& {\tilde G}_{e^{-2 \omega T}}\bigg(
2\omega \, {\rm cotanh}(\omega T)
\bigg({1-\beta^2\over 
1+\beta^2} \bigg)^2 L_0,2\omega \, {\rm cotanh}(\omega T)
\bigg({1-\beta^2\over 
1+\beta^2} \bigg)^2 L_T\bigg),\cr} }
which is nothing but \lutfinlop.

The self-similar kernels under RG transformations have been studied 
and classified in ref. \SPO\ in the context of 1+1-dimensional critical
wetting, and correspond precisely to the form \newparam\ above.   
Each universality class is entirely characterized by the value of
$\mu-1=\beta/(1+\beta^2)$. 
Our original model for decorated Lorentzian triangulations provides
therefore an explicit discrete and simple realization of these universality
classes.

\subsec{Calogero vs random walks}
 
For $\beta<1$ where a continuum limit exists our generalized Lorentzian
triangulation model is equivalent to a one-dimensional Calogero model. 
This model
is known to appear in many different physical problems. 
It would be
interesting to understand whether our generalized Lorentzian triangulation model
thus has an interpretation in terms of a completely different physical
system. 
One particular realization of such a different equivalent physical system
has been found in \FGK\ where the pure Lorentzian case $\beta=0$
was shown to be equivalent to a one-dimensional Random Walk (RW) confined
in a segment of size $t$ upon identifying the fugacity per triangle
$g$ with the fugacity per step of walk. 
The RW model is known to be expressible as
a massive scalar field theory in one dimension with action
\eqn\freefield{
{\cal S}(\varphi)= {1\over 2} \int du \big(\varphi'(u)^2 
+\omega^2 \varphi(u)^2\big),}
which is nothing but the 
ordinary quantum theory of a harmonic oscillator. The quantity 
$\varphi(u)^2$ is interpreted as the density of RW at point $u$,
which precisely translates into the quantity ${\cal N}(u)$ \scalz\
under the abovementioned equivalence, while $\omega$ is the continuous
counterpart of the fugacity per step of walk. 

Surprisingly enough, we have stumbled here on the one-dimensional Calogero
model at $A=1$, rather than the harmonic oscillator corresponding to the action
\freefield\ above. Note that due to the extra potential $1/(8\varphi^2)$,
a crucial difference is that the range of $\varphi$ is over $\IR_+$ 
instead of $\IR$ for the harmonic case. 

\fig{Random Walk picture of pure Lorentzian triangulations. 
A Lorentzian triangulation (a) of time size $t$ and its representation 
as a RW (b) between two reflecting walls at distance $t+1$ (to go from
(b) to (a), simply squeeze the triangles horizontally). The reflected walk 
can be unraveled (c), each portion between two contacts being reflected
or not (here we made for instance two reflections at positions indicated 
by the dashed vertical lines. We finally compactify the time direction
onto a circle of perimeter $2(t+1)$, ending with a RW (d) attached at the 
antipodal points $0$ and $t+1$.}{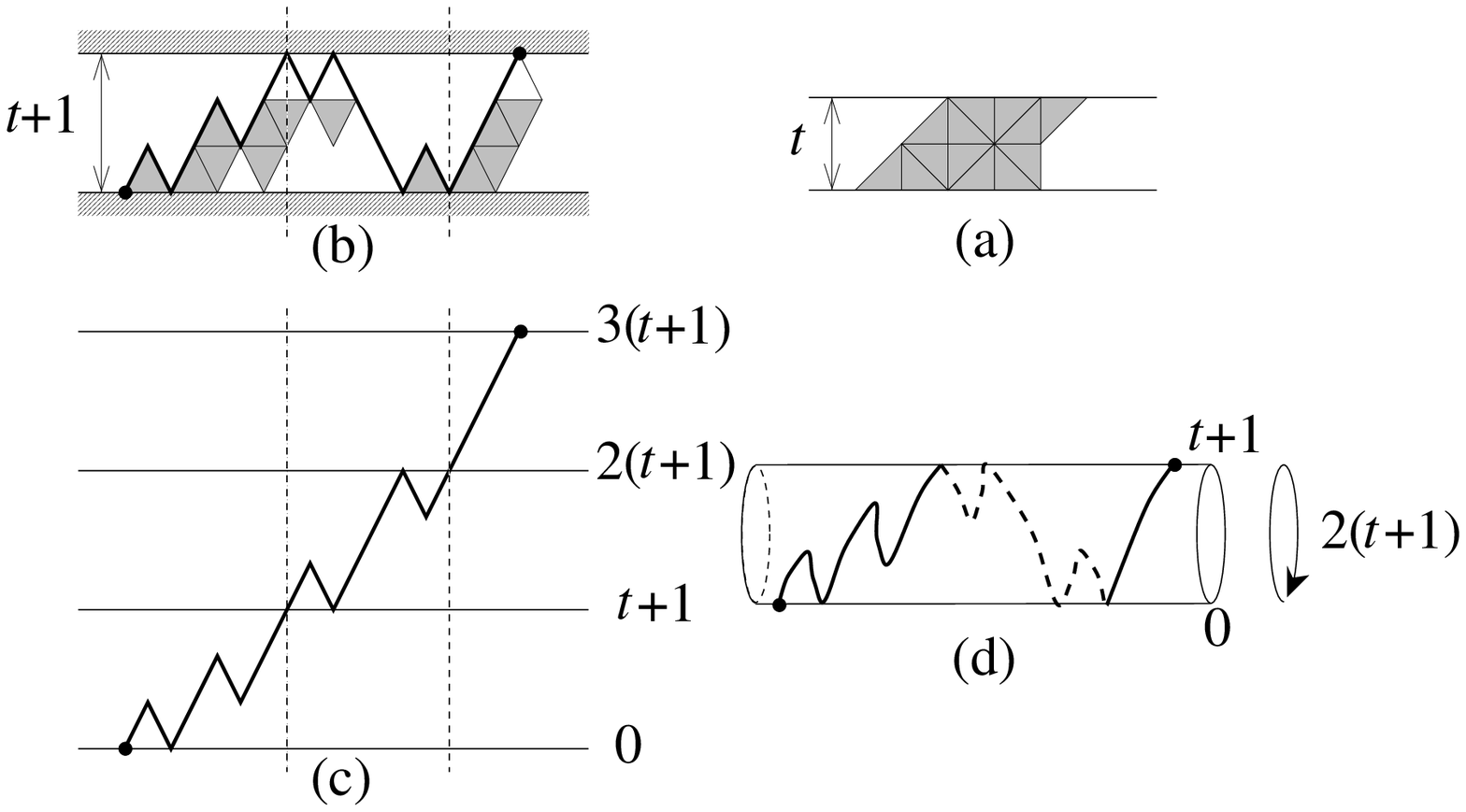}{12.5cm}
\figlabel\wrap

It is an interesting exercise to re-derive the pure Lorentzian triangulation
results
in the RW harmonic oscillator language. In the following we will concentrate
on the loop-loop correlator $Z(i,j,t)$ defined in \geneloop. 
In ref. \FGK, it was shown that 
\eqn\probaz{ Z_{i,j}(t)\equiv g^t Z(i-1,j-1,t)={1\over (2g)^{i+j-1}}
\sum_{S\geq 0} (2g)^S P(i,j,S;t+1) }
where $P(i,j,S;t+1)$ is the probability for a walker on the discrete integer
segment $[0,t+1]\subset \IN$ 
making steps of $\pm 1$ with probability $1/2$ at any integer position 
$0<h<t+1$ and being reflected with probability $1$ whenever it reaches 
the boundaries $h=0$ 
or $t+1$, to go from the position $h=0$ to the position $h=t+1$ in $S$ steps,
with exactly $i$ visits at $h=0$ and $j$ visits at $h=t+1$ (see Fig. \wrap).   
The slightly different transfer matrix $g^{i+j-1}{i+j-2 \choose i-1}$
used in \FGK\ is responsible for
the the shifts by $-1$ of $i$ and $j$, and for the factor $g$ per time step.

The reflection condition at $h=0$ and $t+1$ may be unraveled by associating 
to each walk on $[0,t+1]$ the set of walks on $\IZ$
obtained by iteratively reflecting
or not each portion between two contacts at the boundaries (see Fig. \wrap) 
These new walks now go from the position $h=0$ to some position $h=(2k+1)(t+1)$
with $k\in \IZ$. The numbers of original contacts $i$ and $j$ with respectively 
the
boundaries $h=0$ and $h=t+1$ translate into the total numbers 
of contacts with respectively the lines $h=2k (t+1)$ and $h=(2k+1)(t+1)$, $k\in 
\IZ$.
By compactifying the target space $\IZ$ onto the circle $\IZ/(2(t+1)\IZ)$, we 
end
up with walks from $h=0$ to $h=t+1$ on a circle of perimeter $2(t+1)$ as
shown in Fig. \wrap.
In this final picture, $i$ and $j$ represent respectively the numbers of visits
of the antipodal positions $h=0$ and $h=t+1$.  

In the continuum limit $t = T/a$, $g={1\over 2}(1-a^2 {\omega^2\over 2})$,
the latter RW's are governed by an action of the form \freefield\ with $u\in 
[0,2T]$
and the cyclic boundary condition $\varphi(u+2T)=\varphi(u)$. In this language
the rescaled numbers of visits $ai$ and $aj$ become respectively 
$\varphi(0)^2$ and $\varphi(T)^2$.
Moreover, it is simpler to compute the rescaled loop-loop correlator
\contprop\ with $\beta=0$, by weighting each visit at $h=0$ and $t+1$ 
respectively by $x$ and $y$
and setting $x=1-aX$, $y=1-aY$ as before. We end up with the equivalence
\eqn\equiz{ Z_T(X,Y) ={\langle 2 \varphi(0)\varphi(T) e^{-X\varphi(0)^2}
e^{-Y\varphi(T)^2}\rangle \over  \langle e^{-X\varphi(0)^2}
e^{-Y\varphi(T)^2}\rangle } }
where $\langle ... \rangle$ stands for the functional average using the
abovementioned action for the harmonic oscillator on a circle of perimeter $2T$.  
In eqn.\equiz, the factors $\varphi(0)$ and $\varphi(T)$ ensure
that the walk starts at $0$ and
ends at $T$ (the factor $2$ relates to our choice of normalization of the 
action).  
The insertions of exponential terms account precisely for the weights $x,y$
per visit at $0$, $t+1$, and the denominator ensures the suppression of 
the contribution of disconnected walks.
It is now a simple exercise in 1D quantum mechanics to evaluate the rhs
of \equiz\ namely, using the harmonic oscillator Hamiltonian
$H_{\rm osc}=-{1\over 2} {d^2\over d \varphi^2} +{1\over 2} \omega^2 \varphi^2$:
\eqn\calcharmo{
Z_T(X,Y)={
\int_{-\infty}^\infty 2\varphi_0 d\varphi_0
\varphi_T d\varphi_T
e^{-(\varphi_0^2 X+\varphi_T^2 Y)}
\bigg(\langle \varphi_T \vert e^{-T H_{\rm osc}}\vert \varphi_0\rangle\bigg)^2
\over \int_{-\infty}^\infty d\varphi_0
d\varphi_T e^{-(\varphi_0^2 X+\varphi_T^2 Y)}
\bigg(\langle \varphi_T \vert e^{-T H_{\rm osc}}\vert 
\varphi_0\rangle\bigg)^2},}
to be compared with the Calogero formulation \qmprop.
Using the diagonalization formula (C.4) of appendix C below, with $z=e^{-\omega 
T}$,
we are left with a ratio of two two-dimensional Gaussian integrals
\eqn\goto{ Z_T(X,Y)={\int_{\IR^2} d^2{\vec{\varphi}}\,  2 \varphi_1\varphi_2
e^{-{1\over 2} {\vec{\varphi}}\cdot M {\vec{\varphi}}} \over
\int_{\IR^2} d^2{\vec{\varphi}} 
e^{-{1\over 2} {\vec{\varphi}}\cdot M {\vec{\varphi}}}} = 2 (M^{-1})_{1,2} }
with the matrix 
\eqn\matriM{ M= 2 \pmatrix{ X+\omega \, {\rm cotanh}(T\omega) & -{\omega \over 
{\rm
sinh}(T\omega)} \cr 
-{\omega \over {\rm sinh}(T\omega)} & Y+\omega \, {\rm cotanh}(T\omega) \cr}, }
leading finally to the previous result \loopure, with $\omega=\sqrt{\Lambda}$.

In view of the above derivation, it is remarkable that the Calogero
formulation of the problem automatically takes care of all the 
subtleties (boundary conditions, suppression of disconnected loops,
etc ...) we encountered in the RW approach. In this regard, the 
Calogero formulation of pure Lorentzian triangulations is better adapted.

\subsec{Calogero vs 2D harmonic oscillator}

The previous section is a manifestation of
the well-known connection between the Calogero Hamiltonian at $\mu=1$ and
the harmonic oscillator. 
As we will see now, 
the Schwinger representation we have developed in this paper provides another
direct and exact link
between the pure model and a {\it pair} of (untwisted/twisted) harmonic 
oscillators. This connection holds even {\it before} taking the continuum limit.

Let us start again from the $\beta=0$, $h=1$ counterpart of \schwin\ in which
we also have incorporated for convenience an entropic factor $g$ per
time slice, namely
\eqn\couzer{ g^t Z^{(0)}(t)=
\prod_{s=1}^t\bigg(\oint_{\cal E}
{d \rho_s \over 2 i \pi \rho_s}\int_0^\infty
d\alpha_s e^{-\alpha_s}\bigg)
\prod_{r=1}^t g e^{g\alpha_r (\rho_r+{1\over \rho_{r-1}})},
\ \ \ \ \ \ \rho_{0}=\rho_t. }
We are about to show that the coordinates $\alpha>0$ and $\rho\in {\cal E}$
may be rephrased into polar coordinates of the real plane, pertaining
themselves to two one-dimensional harmonic oscillators.
Performing the change of variable 
\eqn\changeof{\eqalign{\rho_r&=\sqrt{\alpha_{r+1}\over \alpha_{r}}
e^{i \chi_{r+1}}, \qquad r=0,1,...,t-1,\cr \alpha_r&= R_r^2 , \qquad 
r=0,1,...,t, \cr}}
with $\alpha_0\equiv \alpha_t$ and
where we have deformed the contours so as to have
all $\chi_r\in[0,2\pi)$, we get
\eqn\getzerfor{ g^t Z^{(0)}(t)=
\prod_{s=1}^t\bigg( \int_0^{2\pi} {d \chi_s \over 2\pi} \int_0^\infty
2 R_s d R_s e^{-R_s^2}\bigg)
\prod_{r=1}^t g e^{2g R_r R_{r-1}\, {\rm cos}(\chi_r)}. }
Finally, let us perform the following changes of variables on the $\chi$'s,
according to the parity of $t$:
\eqn\parit{\eqalign{
t \ \ {\rm odd}\ &: \ \ \chi_r=\left\{\matrix{
\theta_r+\theta_{r-1},& r=2,3,...,t \cr  
\theta_1+\theta_t, & r=1 \cr}\right.  \cr
t \ \ {\rm even}\ &: \ \ \chi_r=\left\{\matrix{
\theta_r+\theta_{r-1}, &  r=2,3,...,t \cr
\theta_1-\theta_t, & r=1 \cr}\right.  \cr}}
with $\theta_r \in [0,2 \pi)$, $r=1,2,...,t$. 
The extra sign in the case $t$ even is ad-hoc to make the change
of variables invertible. 
We can now interpret 
\eqn\interxy{(x_r = R_r\, {\rm cos}\, \theta_r , 
y_r = R_r\, {\rm sin}\, \theta_r),}
as two real coordinates of the plane $\IR^2$, so that \getzerfor\ becomes
\eqn\getzerfin{g^t Z^{(0)}(t)=\prod_{s=1}^t\bigg({1\over \pi}
\int_{\IR^2} dx_s dy_s e^{-(x_s^2+y_s^2)}\bigg) 
g e^{2g (x_{1}x_t - (-1)^{t-1}y_{1}y_t)}
\prod_{r=2}^t g e^{2g (x_{r-1}x_r - y_{r-1}y_r)} .}
This last formula suggests to introduce the following transfer kernel:
\eqn\tranosc{ G_g(x,y;x',y')= {g\over \pi} \ e^{-{1\over 2}(x^2+y^2+x'^2+y'^2)}
e^{2 g (xx'-yy')}, }
in terms of which the partition function reads:
\eqn\parnewG{ g^t Z^{(0)}(t)=\prod_{s=1}^t\big(
\int_{\IR^2} dx_s dy_s \big) G_g(x_t,(-1)^{t-1}y_t;x_1,y_1)
\prod_{r=2}^t G_g(x_{r-1},y_{r-1};x_r,y_r).}
The transfer kernel \tranosc\ factorizes into $G_g(x,y;x',y')=
G_g^+(x,x')G_g^-(y,y')$, where the two kernels $G_g^\pm$ respectively
correspond to the harmonic oscillator and its twisted version, namely
\eqn\kerplumin{\eqalign{
G_g^+(x,x')&={\sqrt{g}\over \sqrt{\pi}}\ e^{-{1\over 2}(x^2+x'^2-4g xx')}=
\sum_{n\geq 0} \psi_n^{\rm osc}(x) \psi_n^{\rm osc}(x') q^{n+{1\over 2}}, \cr
G_g^-(y,y')&={\sqrt{g}\over \sqrt{\pi}}\ e^{-{1\over 2}(y^2+y'^2+4g yy')}=
\sum_{n\geq 0} (-1)^n \psi_n^{\rm osc}(y) \psi_n^{\rm osc}(y') 
q^{n+{1\over 2}}, \cr }}
where we have set 
\eqn\defq{ g={1\over q+{1\over q}}, }
and the $n$-th eigenfunction of the harmonic oscillator reads
\eqn\harmoeigen{
\psi_n^{\rm osc}(\varphi)= \sqrt{ \sqrt{\omega}
\over \sqrt{\pi} 2^n n!}
H_n(\sqrt{\omega}\varphi) e^{-{1\over 2} \omega \varphi^2},}
where $H_n(x)$ are the Hermite polynomials.
The factorization of the kernel implies that of the partition function
\eqn\partfact{ g^t Z^{(0)}(t)= Z_+^{\rm osc}(t) \ Z_-^{\rm osc}(t),}
where, writing formally products of kernels like ordinary products,
we have
\eqn\partplusmoins{ Z_+^{\rm osc}(t)={\rm Tr}\big((G_g^+)^t\big)={q^{t\over
2}\over 1-q^t}, \qquad
Z_-^{\rm osc}(t)={\rm Tr}\big((G_g^-)^tS^{1+(-1)^t \over 2}\big)={q^{t\over
2}\over 1+q^t},}
where $S$ has the kernel $S(y,y')=\delta(y+y')$. Substituting these into
\partfact\ yields \Zzero.
When we take the continuum limit $a\to 0$ \scalpure,  
both $Z_+^{\rm osc}$ and $Z_-^{\rm osc}$ become functional integrals 
of a scalar free field with action \freefield, the only difference being
the boundary condition, namely $\varphi(u+T)=\varphi(u)$ for the 
oscillator and $\varphi(u+T)=-\varphi(u)$ for its twisted version. Indeed,
in the latter case, we may perform the change of variables $y_{2r}\to -y_{2r}$
for the even indices, while the $y_{2r+1}$ remain unchanged in order to recover
the correct sign of the untwisted case, but for both parities of
$t$ we end up with  a wrong sign for the term $y_1y_t$, translating in the 
continuum limit into the above twisted boundary condition. 
The occurrence of both (twisted and untwisted) sectors arises from the
fact that our boundary 
condition simply imposes that the rescaled number of triangles
$\cal N$ at $u=0$
and $u=T$ are identical, namely $\varphi(u+T)^2=\varphi(u)^2$, which decomposes
into periodic and anti-periodic boundary conditions for the free field itself. 
This factorization property extends to correlations involving the
observables $\Sigma_s(z)$ \factomo, namely
\eqn\forgen{\eqalign{  
\langle \prod_{m=1}^k \Sigma_{s_m}(z_m) \rangle_{\Theta} &=
\langle \prod_{m=1}^k {e^{{\alpha_{s_m} z_m\over 1+z_m}} \over 1+z_m}
\rangle_G \cr
&=\langle \prod_{m=1}^k {e^{{x_{s_m}^2 z_m\over 1+z_m}} \over \sqrt{1+z_m}}
\rangle_{G_g^{+}}\times 
\langle \prod_{m=1}^k {e^{{y_{s_m}^2 z_m\over 1+z_m}} \over \sqrt{1+z_m}}
\rangle_{G_g^{-}}.\cr}}
This explains in particular the factorized form of the one-point
average \fincorzer.
Note that the above steps cannot be retraced for $\beta>0$, where no such
free field formulation seems to exist.

It is instructive to re-derive the loop-loop correlator along the same lines
as above. We start from
\eqn\loptwo{
g^{t+1} Z^{(0)}(x,y,t)=
\prod_{s=1}^{t}\oint_{\cal E}
{d \rho_s \over 2 i \pi \rho_s}
\prod_{s=1}^{t+1}
\int_0^\infty
d\alpha_s e^{-\alpha_s}
\prod_{r=1}^{t+1} g e^{g\alpha_r (\rho_r+{1\over \rho_{r-1}})},
\ \ {1\over \rho_{0}}=x,\, \rho_{t+1}=y. }
Let us change variables as before to $\alpha_r=R_r^2$, $r=1,2,...,t+1$, 
$\rho_r=e^{i\chi_{r+1}}R_{r+1}/R_r$, $r=1,2,...,t$, and 
$\chi_r=\theta_r+\theta_{r-1}$, $r=2,3,...,t+1$, where we choose $\theta_1$
arbitrarily. We end up with $t+1$ two-dimensional integrals over
$(x_r,y_r)=(R_r\, {\rm cos}\,\theta_r,R_r\, {\rm sin}\,\theta_r)$,
$r=1,2,...,t+1$, by adding for convenience
an extra integral ${1\over 2\pi}\int_0^{2\pi} d\theta_1$, since 
the result does not depend on the choice $\theta_1$. 
This yields
\eqn\xyint{\eqalign{
g^{t+1} &Z^{(0)}(x,y,t)= \cr 
&\prod_{s=1}^{t+1}\big( {1\over \pi}
\int_{\IR^2} dx_s dy_s  
e^{-(x_s^2+y_s^2)}\bigg)
g e^{gx(x_1^2+y_1^2)+gy(x_{t+1}^2+y_{t+1}^2)}
\prod_{r=2}^{t+1} g e^{2g (x_{r-1}x_r - y_{r-1}y_r)} .\cr}}
Note that if we switch to $y_{2r}\to y_{2r}$ and 
$y_{2r-1}\to -y_{2r-1}$, the integrals over the $x$'s and the $y$'s are
identical, leading to
\eqn\deterxy{\eqalign{
g^{t+1} Z^{(0)}(x,y,t)&=
{\rm det}^{-1}\pmatrix{ {1\over g}-x & -1 & 0 & \cdots & 0 \cr
-1 & {1\over g}& -1 & \ddots & \vdots \cr
0 & \ddots & \ddots & \ddots & 0 \cr
\vdots &  \ddots & -1 & {1\over g} & -1 \cr
0 & \cdots &    0 & -1 & {1\over g}-y  \cr} \cr
&= {q^t (1-q^2) \over 
(1-qx)(1-qy) -q^{2t} (q-x)(q-y)}, \cr}}
where ${1\over g}=q+{1\over q}$, and the matrix is of size $(t+1)\times (t+1)$.
This is in perfect agreement with the result of ref. \FGK\ and with the 
continuum limit \loopure.

To conclude this section, we have unearthed an explicit two-component
free field structure in the pure Lorentzian triangulation model, as opposed
to the one-component free field description of Sect.\ 6.3 above.
This increase in dimension is apparently the price to pay for reducing 
the interval $[0,2T]$ of the latter description to the original one $[0,T]$.
Indeed the information needed to reconstruct the observables
at time $u$ pertains to the {\it pair} of slices $u$ and $2T-u$ 
in the one-dimensional free field model.

This two-dimensional structure 
{\it a posteriori} explains the emergence of the $A=1$ Calogero model in its 
continuum description: indeed, the corresponding two-dimensional
Schr\"odinger equation 
\eqn\scrtwo{ \bigg(-{1\over 2}( {\partial_x^2 }+{\partial_y^2})
+{1\over 2}\omega^2 (x^2+y^2) \bigg) \psi(x,y\vert t)= \partial_t 
\psi(x,y\vert t), } 
when restricted to the rotation-invariant sector simply reads
\eqn\newschro{ \bigg(-{1\over 2}( \partial_R^2 +{1\over R}\partial_R )
+{1\over 2}\omega^2 R^2 \bigg) \psi(R\vert t) =\partial_t \psi(R\vert t). }
Upon redefining $\psi(R\vert t)= R^{-{1\over 2}} \phi(R\vert t)$, we end
up with the Schr\"odinger equation with Calogero potential 
\eqn\caloq{ \bigg(-{1\over 2}\partial_R^2 +{1\over 2}\omega^2 R^2 
-{1\over 8 R^2} \bigg) \phi(R\vert t)=\partial_t \phi(R\vert t).}
In this section, we have only considered radial observables thus ignoring
completely the angular sector of the theory: it would be interesting to 
find the meaning of the angles in the triangulation language.

Finally, it would be interesting to understand if it is 
possible to construct a surface (or volume) model for which the continuum
theory would belong to the class of higher dimensional (multi-particle) 
Calogero models which are known to have a much richer structure 
than the simple one, but this is another story.

\vskip 0.7cm

\noindent
{\bf Acknowledgments: }

\noindent
We thank J.\ Ambj\o rn, M. Berg\`ere and T. Garel for useful discussions.
All authors acknowledge support by the EU network on ``Discrete
Random Geometry", grant HPRN-CT-1999-00161.

\appendix{A}{An alternative analysis of the transfer matrix}

In this section we show how to recover the Calogero Hamiltonian with $A=1$
in the pure case using the strategy of ref.~\AL. 
We also 
explain why the same strategy becomes much more involved for $\beta\neq 0$
and cannot apply to $\beta\geq 1$. 

The starting point is the composition law~\AL\
\eqn\contour{
\Theta^{(1)}(x,y,t+1)=\oint_{\cal C}{d\omega \over 2 \pi i \omega}
\Theta^{(1)}\left(x,{1 \over \omega}\right) \Theta^{(1)}(\omega,y,t),}
where $\Theta^{(1)}(x,y,t)=\left(\Theta^{(1)}\right)^t(x,y)$ 
%of eqn. \geneloop\ at $\beta=0$, 
with $\Theta^{(1)}(x,y)$  given by~\thetas. 
Here we can choose the contour ${\cal C}$ to encircle the singularities of 
${1 \over \omega}\Theta^{(1)}\left(x,{1\over \omega}\right)$, which consist
of a simple pole, and not those of $\Theta^{(1)}(\omega,y,t)$.
Inserting the
scaling relations
\eqn\scalvar{\eqalign{
g={1 \over 2}\left(1-{1\over 2}a^2\, \Lambda\right),
\ \ &x=1-a X,\ \
y=1-a Y,\ \ \omega=1-aZ, \ \ t={T\over a}, \cr
&\Theta^{(1)}(X,Y,T)\equiv {a \over 2^t}\Theta^{(1)}(x,y,t) , \cr}}
and assuming the time evolution to be described in the scaling limit by
a Hamiltonian $\hat{H}$, one gets for the evolution on an elementary time lapse
$T=a$:
\eqn\Hamexp{\eqalign{
(1-a \hat{H}+O(a^2))\psi& ={1 \over 2}\oint {d\omega \over 2 \pi i \omega} 
\Theta^{(1)}\left(x,{1 \over \omega}\right) \psi(\omega) \cr
%&=\int_{-i\infty+c}^{i\infty+c} {dZ \over 2 \pi i}
%{Z_a^{(0)}(X,-Z)\over 1-aZ} \psi(Z) \cr
&= \int_{-i\infty+c}^{i\infty+c} {dZ \over 2 \pi i} 
\left\{{1 \over Z-X}+a\left({Z\over Z-X}+{\Lambda -Z^2\over
(Z-X)^2}\right)+O(a^2)\right\} \psi(Z),}} 
where $c$ is chosen so that
the integration contour lies to the right of the pole at $Z=X$, and we have
denoted by the same letter $\psi(\omega=1-aZ)\equiv \psi(Z)$.
This gives
\eqn\Hamx{
\hat{H}=(X^2-\Lambda){\partial \over \partial X}+X.}
Writing $\psi(X)=\int_0^{\infty}dL e^{-XL}\psi(L)$, $L$ has the interpretation 
of
the continuum loop length (cf.\ eqn.~\Laplace) and the inverse-
Laplace transformed Hamiltonian $\hat{H}$ acting on functions of $L$
takes the form
\eqn\HamL{
\hat{H}=-L{\partial^2 \over \partial L^2}-{\partial \over \partial L}+\Lambda 
L,}
with a flat measure on $L$ for the wave functions $\psi(L)$.
Then performing the simultaneous change of variables and functions
\eqn\autoroute{
L={1\over 2}\varphi^2, \qquad 
\phi(\varphi)=\sqrt{\varphi}\psi\left({\varphi^2\over
2}\right), }
guaranteeing a flat measure on $\varphi$ for $\phi(\varphi)$, we
finally get
\eqn\Hamphi{
\hat{H}=-{1 \over 2} {\partial^2 \over \partial \varphi^2} +{1\over 2} \Lambda
\varphi^2-{1\over 8}{1\over \varphi^2},}
which coincides exactly with the result~\hamilto\  for $\beta=0$.

Now, let us try to apply the same strategy for $\beta\neq 0$. Evidently, we
need first to evaluate the transfer matrix. For the present calculation it
proves convenient to consider its symmetrized version, i.e.\
\eqn\thetacom{
\Theta(x,y)=\oint_{\cal C} {d\omega_1 \over 2 \pi i\omega_1} \oint_{\cal C} 
{d\omega_2 \over 2 \pi
i\omega_2} (\Theta^{(1)})^{1/2}\left(x,{1\over \omega_1}\right)
\Theta^{(2)}(\omega_1,\omega_2) (\Theta^{(1)})^{1/2}\left({1 \over\omega_2},y
\right).}
Since the transfer matrix of the pure model has been explicitly
diagonalized~\FGK, it is straightforward to write down an expression for
$(\Theta^{(1)})^{1/2}(x,y)$
\eqn\kvadratrod{
(\Theta^{(1)})^{1/2}(x,y)={C_1(x) \over 1-C_2(x) y}={C_1(y) \over 1-C_2(y)x},} 
where 
\eqn\cfunc{
C_1(x)={(1+2g)^{1/2}\over 1+g(1-x)}, \ \ \ \ \ \
C_2(x)={g(1+x)\over 1+g(1-x)}.}
Here we have chosen to write $(\Theta^{(1)})^{1/2}(x,y)$ in a form which
explicitly exposes its singularity structure. Notice that 
$(\Theta^{(1)})^{1/2}(x,y)$
is symmetric in $x$ and $y$ as it should be. Inserting the 
expression~\kvadratrod\ for
$(\Theta^{(1)})^{1/2}$ and picking the pole $\omega_1=C_2(x)$ in the first term
and $\omega_2=C_2(y)$ in the last one, we end up with
\eqn\final{\eqalign{
\Theta(x,y)&=
C_1(x)C_1(y) \Theta^{(2)}(C_2(x),C_2(y)) \cr
&={C_1(x)C_1(y) \over
  \sqrt{1-4\theta^2(C_2(x))^2}\sqrt{1-4\theta^2(C_2(y))^2}} 
{1\over 1-{4{\theta^2 \over \beta^2} 
C_2(x)C_2(y)\over  \left(1+  \sqrt{1-4\theta^2(C_2(x))^2}\right)
\left(1+\sqrt{1-4\theta^2(C_2(y))^2}\right)}}. \cr}}
{}From this expression
for the transfer matrix we can read off various facts about the critical
properties of our model. First, we see that there are several ways in which
the transfer matrix can become singular. One possibility is that the argument
of the square root vanishes. Assuming that as usual the critical values\foot{
The reader should not be surprised that the critical values of $x$ and $y$
differ from those of \scanz. Indeed,
the critical values $w_c=z_c={1\over 2g}$
of \scanz\ correspond to the vanishing of the denominator of
$\Theta^{(1)}(w,z)=\oint {dx \over 2i\pi x} {dy \over 2i\pi y} 
(\Theta^{(1)})^{1/2}(w,y) {1\over 1-xy}(\Theta^{(1)})^{1/2}(x,z)$, 
while the denominator of
$(\Theta^{(1)})^{1/2}(x,z)$ vanishes for $x=x_c=1$ when $z=z_c$, and similarly 
for
$y=y_c=1$ when $w=w_c$ (by use of \cfunc).}
of $x$ and $y$ are $x_c=y_c=1$ this situation occurs when 
\eqn\gsing{
4g\theta=1.}
Another possibility is that the denominator of the second factor in~\final\
vanishes. This corresponds to the situation
\eqn\betasing{
2g\theta\left(\beta+{1\over \beta}\right)=1, \ \ \ \ \ \ \ {\rm and}
\ \ \ \ \ \beta\leq 1.}
The critical point \betasing\ is the one
which reduces to the pure Lorentzian triangulation critical point 
when $\beta\to 0$ (more
precisely when $h=1$ and $\theta\to 0$). The singularity given by~\gsing\
corresponds to the critical point of an isolated arch system. As long as
$\beta<1$ we always reach the critical point of \betasing\ 
before that of \gsing. For $\beta=1$ the two points coincide and 
finally for $\beta>1$ only the singularity \gsing\ persists. This
analysis matches exactly our results in Sect. 4, in particular the
relations~\singone, \singb\ and \critipt.

It is obvious that the scaling behavior of $\Theta(x,y)$ in the vicinity of
a singularity corresponding to the vanishing of the square root is not
compatible with the existence of a Hamiltonian (cf.\ relation~\Hamexp).
Thus we can only hope to be able to introduce a meaningful continuum time
variable as long as $\beta<1$. This conclusion is of course in agreement
with the analysis of Sect. 4. It is, however, not
straightforward to write down the Hamiltonian for $\beta<1$ using the
relations~\contour\ and~\Hamexp, the reason being that for the generalized
model we will get not only a pole contribution to the contour integral 
in~\Hamexp\ but also a cut contribution.
Introducing the scaling relations
\eqn\scalcomb{
g={1\over 2} \left(1-{1\over 2}a^2 \Lambda \right),
\ \ \ \ \ \ x=1-aX,\ \ \ \ \ \ \theta=\left(\beta+{1\over \beta}\right)^{-1},}
the contribution from the pole in~\final\ reproduces the 
the same structure as that of \Hamexp, up to a redefinition of the 
renormalized fugacity per triangle,
$\Lambda \to \tilde{\Lambda}=\Lambda 
\left({1+\beta^2 \over 1-\beta^2}\right)^2$.
The contribution from the cut is more involved and cannot be obtained so
simply. 
This is not surprising: the Hamiltonian \hamilto\ (using \freq)
should read in the language of
modified lengths ${\tilde L}={1\over 2}\varphi^2=\left({1-\beta^2\over 
1+\beta^2}
\right)^2 L$:
\eqn\Heqal{ {\hat H } = -{\tilde L}{\partial^2 \over \partial {\tilde L}^2} 
-{\partial \over \partial {\tilde L}} +\omega^2 {\tilde L}  
+{(\mu-1)^2\over 4 {\tilde L}}.} 
When Laplace-transformed in terms of $\tilde X$, it becomes
\eqn\becoH{ {\hat H}\, \psi({\tilde X}) = \bigg( ({\tilde X}^2 -\omega^2) 
{\partial\over 
\partial {\tilde X}} + {\tilde X}\bigg) \psi({\tilde X}) +{(\mu-1)^2 \over 4}
 \int_{\tilde X}^\infty
\psi(u) du ,}
showing that the cut contribution must give rise to a non-local integral term.

\appendix{B}{The $k$-point function of the scaled numbers of triangles for
$\beta <1$}

In terms of the time intervals
$t_m=u_{m+1}-u_m$ for $m=1,2,...k-1$, and $t_k=T-\sum_{1\leq m \leq k-1} t_m$
the $k$-point correlator
of $e^{{\cal Z}_m\omega {\cal N}(u_m)}$ reads
\eqn\readbet{\eqalign{
\langle \prod_{m=1}^k &e^{{\cal Z}_m\omega {\cal N}(u_m)}\rangle=
\langle \prod_{m=1}^k e^{{\cal Z}_m\omega\varphi^2(u_m)} \rangle\cr
&={1\over Z_{T,\mu}}\int_0^\infty d\varphi_1...d\varphi_k
\sum_{n_1,...,n_k\geq 0} \prod_{i=1}^k
e^{\omega {\cal Z}_i\varphi^2_i}
\psi_{n_i}(\varphi_i)\psi_{n_i}(\varphi_{i+1}) 
e^{-\omega t_i(2n_i+\mu)}, \cr }}
where we have defined $\varphi_{k+1}\equiv\varphi_1$.
We now use $k$ times the quadratic relation for the modified Laguerre 
polynomials
\laquad\
and the change of variables $v_i=\omega\varphi_i^2$, $i=1,2,...,k$
to rewrite
\eqn\calculpenible{\langle \prod_{m=1}^k e^{{\cal Z}_m\omega {\cal
N}(u_m)}\rangle= 2\, {\rm sinh}(\omega T) e^{\omega T(\mu-1)} 
\int_0^\infty \prod_{i=1}^k x_i dv_i \prod_{i=1}^k e^{-\delta_i v_i}
I_{\mu-1}(2x_i\sqrt{v_iv_{i+1}}) ,}
where we have set
\eqn\setxiga{
\eqalign{x_i &={1\over 2\, {\rm sinh}(\omega t_i)}, \qquad 
\delta_i=\gamma_i-{\cal Z}_i ,\cr
\gamma_i&= 2 x_i x_{i-1}\, {\rm sinh}(\omega (t_i+t_{i-1})) , \cr}}
with the convention that $t_0\equiv t_k$.
Let us evaluate by induction the integral
\eqn\inducJ{
J_k(x_1,...,x_k\vert \delta_1,...,\delta_k)=2{\rm sinh}(\omega T)
e^{\omega T(\mu-1)} \int_0^\infty
\prod_{i=1}^k x_i dv_i
\prod_{i=1}^k e^{-\delta_i v_i}I_{\mu-1}(2x_i\sqrt{v_{i}v_{i+1}}) .}
We first perform the integration over
$v_k$ by applying the following integral relation
\eqn\usefulagain{ r \int_0^\infty dv \, e^{-r v} I_{\mu-1}(2 s \sqrt{v})
I_{\mu-1}(2 s' \sqrt{v})= e^{{1\over r}(s^2+s'^2)} \, I_{\mu-1}\left({2 s 
s'\over 
r}\right), }
with $r=\delta_k$,
$s=x_{k-1}\sqrt{v_{k-1}}$ and $s'=x_k\sqrt{v_1}$, with the result
\eqn\resulJ{
J_k(x_1,...,x_k\vert \delta_1,...,\delta_k)=
J_{k-1}(x_1,x_2,...,{x_{k-1}x_k\over \delta_k}\vert
\delta_1-{x_k^2\over \delta_k},\delta_2,...,\delta_{k-2},
\delta_{k-1}-{x_{k-1}^2\over \delta_k}) .}
This is valid for all $k\geq 3$. When $k=2$, we simply get
a single integral
\eqn\singlinteg{
J_2(x_1,x_2\vert\delta_1,\delta_2)={\rm sinh}(\omega T) e^{\omega T(\mu-1)}
{2x_1x_2\over \delta_2}\int_0^\infty dv
e^{-(\delta_1 -{x_1^2+x_2^2\over \delta_2})v} I_{\mu-1}(2 v x_1x_2/\delta_2),}
readily evaluated as a particular case of the Laplace transform of the modified
Bessel function \useful\ with $r=\delta_1 -{x_1^2+x_2^2\over \delta_2}$ and
$s=2 x_1x_2/\delta_2$. 
This gives
\eqn\givesJ{
\eqalign{J_2(x_1,x_2\vert\delta_1,\delta_2)&=g_\mu\left({2 x_1 x_2 \over
\delta_1\delta_2-x_1^2-x_2^2}\right), \cr
g_\mu(s)&={\rm sinh}(\omega T) e^{\omega T(\mu-1)}
s^{2-\mu} {(1-\sqrt{1-s^2})^{\mu-1}\over \sqrt{1-s^2}} .\cr}}
Similarly, for $k=1$ we get
\eqn\Jone{\langle e^{{\cal Z}\omega {\cal N}}\rangle=
f_{\mu}({\cal Z})=J_1(x_1\vert \delta_1)= g_\mu({2 x_1 \over \delta_1})=
g_\mu({1\over {\rm cosh}(\omega T)-{\cal Z} {\rm sinh}(\omega T)}),}
In general, applying the recursion relation \resulJ, we find
\eqn\generecJ{
\langle \prod_{m=1}^k e^{\omega {\cal N}(u_m) {\cal Z}_m}\rangle
=g_\mu\big({2 \over P({\cal Z}_1,...,{\cal Z}_k \vert t_1,...,t_k)}\big),}
where the polynomial $P$ is defined as follows:
we first introduce the sequence
\eqn\seqmu{
\mu_{k-p}=\delta_{k-p}-{x_{k-p}^2\over \mu_{k-p+1}}, \ \ p=1,2,...,k-1 \ \
{\rm with} \qquad \mu_k=\delta_k, }
and the polynomial $P$ reads
\eqn\finPJ{
P({\cal Z}_1,{\cal Z}_2,...,{\cal Z}_k\vert t_1,t_2,...,t_k)= 
{\mu_1 \mu_2 ... \mu_k\over x_1 x_2 ...x_k}-
\sum_{p=0}^{k-2} {\mu_2 \mu_3 ... \mu_{k-p}\over x_1 x_2 ...x_{k-p-1}}\times
{x_{k-p} x_{k-p+1}... x_k \over \mu_{k-p} \mu_{k-p+1} ... \mu_k}.}
Once expressed in terms of ${\cal Z}_i$ and $t_i$, this polynomial takes the 
very
simple form \Ppol, with for instance
\eqn\qqps{\eqalign{
&k=0: \ \ \ P=2 {\rm cosh}(\omega T), \cr
&k=1: \ \ \ P({\cal Z}_1\vert t_1=T)=2\big( {\rm cosh}(\omega T)- 
{\cal Z}_1 {\rm sinh}(\omega T) \big),\cr
&k=2: \ \ \ P({\cal Z}_1,{\cal Z}_2\vert t_1,t_2=T-t_1)=
2\big( {\rm cosh}(\omega T)-
({\cal Z}_1+{\cal Z}_2) {\rm sinh}(\omega T)\cr
&\ \ \ \ \ \ \ \ \ \ \ +2{\cal Z}_1 {\cal Z}_2
{\rm sinh}(\omega t_1) {\rm sinh}(\omega t_2) \big),\cr}}
leading respectively to the correlators
\eqn\zeronetwo{\eqalign{
&k=0: \ \ \ \langle 1 \rangle =g_\mu({2\over P})=
f_\mu(0)=1, \cr
&k=1: \ \ \ \langle e^{\omega {\cal N} {\cal Z}}\rangle=g_\mu\big(
{2\over P({\cal Z}\vert T)}\big)=f_\mu({\cal Z}), \cr
&k=2: \ \ \ \langle e^{\omega {\cal N}(u_1) {\cal Z}_1}
e^{\omega {\cal N}(u_2) {\cal Z}_2}\rangle= g_\mu\bigg(
{2\over P({\cal Z}_1,{\cal Z}_2\vert u_2-u_1,T+u_1-u_2)}\bigg)\cr
&\ \ \ \ \ \ \ \ \ \ \ \ \ \ \ \ \ \ \ \ \ =
f_\mu\left({\cal Z}_1+{\cal Z}_2-2{\cal Z}_1{\cal Z}_2{{\rm 
sinh}(\omega(u_2-u_1))
{\rm sinh}(\omega(T+u_1-u_2)) \over {\rm sinh}(\omega T)}\right),\cr}}
with $f_\mu$ as in \oneptfin. More generally, the $k$-point correlator is
expressed as 
\eqn\gmufmu{g_\mu\big(s={2\over P_k}\big)=
f_\mu\big({\rm cotanh}(\omega T)-{P_k\over 2\, {\rm sinh}(\omega T)}\big), }
in terms of $P_k\equiv P({\cal Z}_1,...,{\cal Z}_k\vert t_1,...,t_k)$.

As a final check on the result \Ppol,
we see that
\eqn\Ppolcheck{\eqalign{
P({\cal Z}_1,...,{\cal Z}_{k-1},{\cal Z}_k=0\vert t_1,...,t_k)
&=P({\cal Z}_1,...,{\cal Z}_{k-1}\vert t_1,...,t_{k-1})\cr
P({\cal Z}_1,...,{\cal Z}_k\vert t_1,...,t_{k-1},t_k=0)
&=P({\cal Z}_1,...,{\cal Z}_{k-2},{\cal Z}_{k-1}+{\cal Z}_k\vert
t_1,...,t_{k-1}),\cr}}
as expected from the definition of the correlator.

\appendix{C}{ Comparison with the harmonic oscillator}

We wish to compute the $k$-point  correlation function
of the operator $e^{\omega\varphi^2(u) {\cal Z}}$ in the
one-dimensional quantum system with Hamiltonian
\eqn\zeroham{ H_{\rm osc}=-{1\over 2}{d^2 \over d\varphi^2} 
+{1\over 2}\omega^2\varphi^2,}
on a time circle of length $T$. The Hamiltonian has eigenvalues 
$E_n=\omega(n+{1\over 2})$ and eigenfunctions
\eqn\eigharmo{ \psi_n^{\rm osc}(\varphi)= \sqrt{ \sqrt{\omega} 
\over \sqrt{\pi} 2^n n!}
H_n(\sqrt{\omega}\varphi) e^{-{1\over 2} \omega \varphi^2},}
where $H_n(x)$ are the Hermite polynomials,
$n=0,1,2,...$ and $\varphi\in (-\infty,+\infty)$. Next we will use
the following quadratic relation obeyed by the  
Hermite polynomials
\eqn\quadherm{\sum_{n=0}^\infty H_n(x)H_n(y) {(z/2)^n \over n!}
={1\over \sqrt{1-z^2}} e^{2xyz-z^2(x^2+y^2)\over 1-z^2}, }
immediately translated into
\eqn\transherm{\sum_{n\geq 0} \psi_n^{\rm osc}(\varphi_0)\psi_n^{\rm osc}
(\varphi_1) z^{n+{1\over 2}}=
\sqrt{\omega z \over \pi (1-z^2)} e^{{2z\over 1-z^2}\omega\varphi_0\varphi_1
-{1+z^2\over 1-z^2}{\omega\over 2}(\varphi_0^2+\varphi_1^2)} .}
The correlation function reads
\eqn\nptherm{\eqalign{
\langle \prod_{m=1}^k &e^{\omega\varphi^2(u_m){\cal Z}_m}\rangle_{\rm osc}\cr
&=
{1\over Z_{T,\rm osc}}\int_\IR d\varphi_1...d\varphi_k \sum_{n_1,...,n_k\geq 0}
\prod_{i=1}^k
e^{-\omega t_i (n_i+{1\over 2})} \psi_{n_i}^{\rm osc}(\varphi_i)
\psi_{n_i}^{\rm osc}(\varphi_{i+1}) e^{{\cal Z}_i\omega \varphi_i^2},\cr}}
where the time intervals $t_i=u_{i+1}-u_i$, $i=1,2,...,k-1$, 
$t_k=T-\sum_{1\leq i\leq k-1} t_i$.
Applying $k$ times the relation \transherm\
with respectively
$\varphi_0,\varphi_1\to \varphi_i,\varphi_{i+1}$ and $z\to e^{-\omega t_i}$,
and performing the change of variables $v_i= \sqrt{2 \omega} \, \varphi_i$, we 
get
\eqn\getherm{\langle \prod_{m=1}^k e^{\omega\varphi^2(u_m){\cal 
Z}_m}\rangle_{\rm
osc}={\sqrt{x_1...x_k}\over Z_{T,{\rm osc}}}\int_\IR \big(\prod_{i=1}^k
{dv_i\over\sqrt{2\pi}} \big) \ e^{-{1\over 2}\sum_{i=1}^k \delta_i v_i^2-2
x_iv_iv_{i+1}}, }
with $x_i$ and $\delta_i$ as in \setxiga.
This finally yields
\eqn\finherm{ \langle \prod_{m=1}^k e^{\omega\varphi^2(u_m){\cal 
Z}_m}\rangle_{\rm
osc}=2 \, {\rm sinh}(\omega T/2) \sqrt{x_1...x_k}\, {\rm det}(M-Z)^{-{1\over 2}}
=\sqrt{ {\rm det}(M) \over {\rm det}(M-Z)} ,}
as a straightforward result of the Gaussian integration, with $M$ and $Z$ as in
\matrimz.
The comparison with the alternative expression \harmoco\ immediately
yields the determinantal expression for the polynomial $P$ \deterexp.

\listrefs

\bye